\documentclass[12pt]{article}

\usepackage[latin1]{inputenc}
\usepackage[usenames,dvipsnames,pdftex]{xcolor}
\usepackage{amsmath,amssymb,amsfonts,mathrsfs}
\usepackage{miller}
\usepackage{subeqnarray}
\usepackage{subcaption}
\usepackage{pdflscape}
\usepackage{verbatim}
\usepackage{lipsum}
\usepackage{bm}
\usepackage{tikz}
\usepackage[margin=2cm]{geometry}
\usetikzlibrary{arrows,shapes,positioning}
\usetikzlibrary{decorations.markings}
\tikzstyle arrowstyle=[scale=1]
\tikzstyle directed=[postaction={decorate,decoration={markings,
    mark=at position 1 with {\arrow[arrowstyle]{stealth}}}}]
\usepackage{textcmds}
\usepackage[super]{nth}
\usetikzlibrary{arrows}
\usepackage{booktabs}
\DeclareCaptionLabelFormat{cont}{#1~#2 (continued)}
\usepackage{siunitx}
\usepackage{dblfloatfix}

\DeclareSymbolFont{bbold}{U}{bbold}{m}{n}
\DeclareSymbolFontAlphabet{\mathbbold}{bbold}
\providecolor[named]{Ref}{rgb}{0 0 0.4}
\usepackage[title]{appendix}
\usepackage{graphicx}
\usepackage{multirow}
\graphicspath{{./figures/}{./}{./fig/}}
\DeclareGraphicsExtensions{.pdf,.png,.jpg}
\usepackage[pdftex,bookmarksnumbered=true,colorlinks=true,allcolors=black,pdfpagelabels=false,plainpages=false]{hyperref}
\usepackage[capitalize]{cleveref}

\newcommand{\term}[1]{\textit{#1}}
\newcommand{\ie}{\textit{i.e.}}
\newcommand{\eg}{\textit{e.g.}}
\newcommand{\vs}{\textit{vs.}}

\newcommand{\transpose}[1]{\ensuremath{{#1}^{\mathrm T}}}

\newcommand{\inverse}[1]{\ensuremath{{#1}^{-1}}}

\newcommand{\doublecon}{\ensuremath{:}}
\newcommand{\invtranspose}[1]{\ensuremath{{#1}^{\mathrm{-T}}}}
\newcommand{\sign}[1]{\ensuremath{\operatorname{sgn}\left({#1}\right)}}

\newcommand{\totalder}[2]{\ensuremath{\frac{\inc{#1}}{\inc{#2}}}}
\newcommand{\partialder}[2]{\ensuremath{\frac{\partial{#1}}{\partial{#2}}}}
\newcommand{\inc}[1]{\ensuremath{\text d{#1}}}
\newcommand{\abs}[1]{\ensuremath{\left|{#1}\right|}}
\newcommand{\norm}[2][]{\ensuremath{\left|\left|{#2}\right|\right|\ifx &#1&\else _{#1}\fi}}

\newcommand{\tnsrfour}[1]{\ensuremath{\mathbb{#1}}}
\newcommand{\tnsr}[1]{\ensuremath{\mathbf{#1}}}

\newcommand{\vctr}[1]{\ensuremath{\mathbf{#1}}}

\newcommand{\eyetwo}{\ensuremath{\tnsr I}}

\newcommand{\fPK}{\ensuremath{\tnsr P}}

\newcommand{\sPK}{\ensuremath{\tnsr S}}
\newcommand{\F}[1][]{\ensuremath{\tnsr F^{#1}}}

\newcommand{\Fp}[1][]{\ensuremath{\tnsr F_\text{p}^{#1}}}

\newcommand{\Fe}[1][]{\ensuremath{\tnsr F_\text{e}^{#1}}}

\newcommand{\R}[1][]{\ensuremath{\tnsr R}}
\newcommand{\U}[1][]{\ensuremath{\tnsr U}}
\newcommand{\V}[1][]{\ensuremath{\tnsr V}}

\newcommand{\velgrad}{\ensuremath{\tnsr L}}
\newcommand{\Lp}{\ensuremath{\tnsr L_\text{p}}}
\newcommand{\Le}{\ensuremath{\tnsr L_\text{e}}}

\newcommand{\Emajor}{\ensuremath{\varepsilon_{\text{major}}}}
\newcommand{\Eminor}{\ensuremath{\varepsilon_{\text{minor}}}}
\title{Assessment of full field crystal plasticity finite element method for forming limit diagram prediction}

\author{Duancheng Ma \\ 
{\normalsize LKR Leichtmetallkompetenzzentrum Ranshofen GmbH,} \\
{\normalsize AIT Austrian Institute of Technology,} \\ 
{\normalsize Lamprechtshausenerstra\ss e 61,} \\
{\normalsize Postfach 26, 5282 Ranshofen-Braunau, Austria}\\
{\normalsize duancheng.ma@ait.ac.at}}
\vspace{10pt}
\date{}

\begin{document}
\maketitle

\begin{abstract}
This study is inspired by the recent development of ``virtual material testing laboratory'' in which the main equipment is a full field crystal plasticity modeling tool. Ample examples have demonstrated its applications to sheet forming operations. In those applications, the mechanical anisotropy originated from the crystallographic texture can be adequately described, such as r-values and earing. Formability is another important property in sheet metal forming, which has yet not been equipped in these virtual laboratories. Though theoretical models for formability can be dated back to 1885 due to Consid\`{e}re, all popular models at the moment suffer from their respective limitations. In this study, we explore the feasibility of applying full field crystal plasticity model for calculating forming limit diagram, avoiding using additional assumptions or models for determining the forming limits.
\end{abstract}
\section{Introduction}
\label{sec:introduction}
\subsection{Forming limit diagram}
\label{sec:intro_fld}
\begin{figure*}[h!]
\begin{subfigure}[b]{0.5\linewidth}
\centering
\includegraphics[scale=0.6]{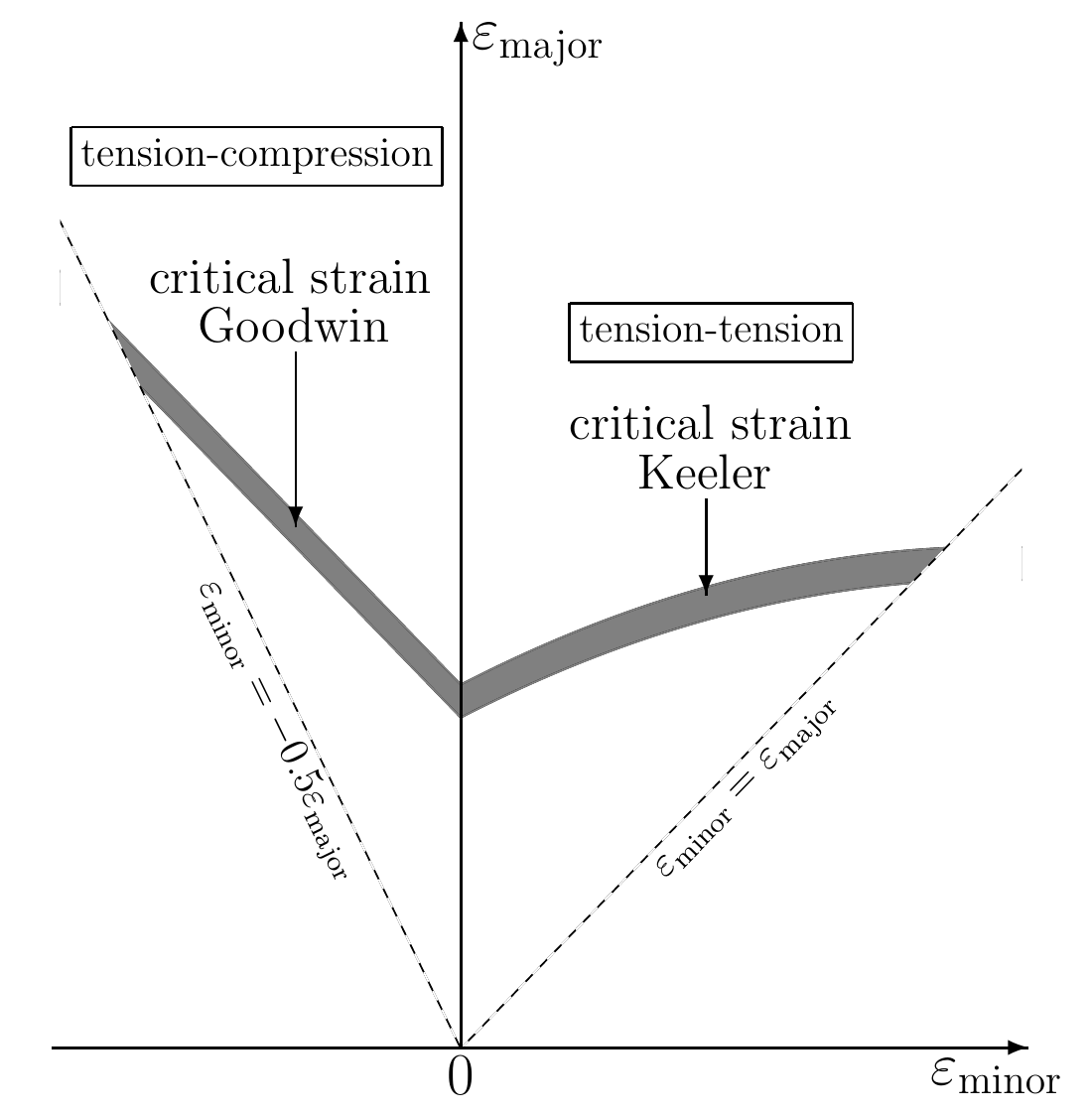}
\subcaption{}
\label{fig:intro_fld_1}
\end{subfigure}
\begin{subfigure}[b]{0.5\linewidth}
\centering
\includegraphics[scale=0.6]{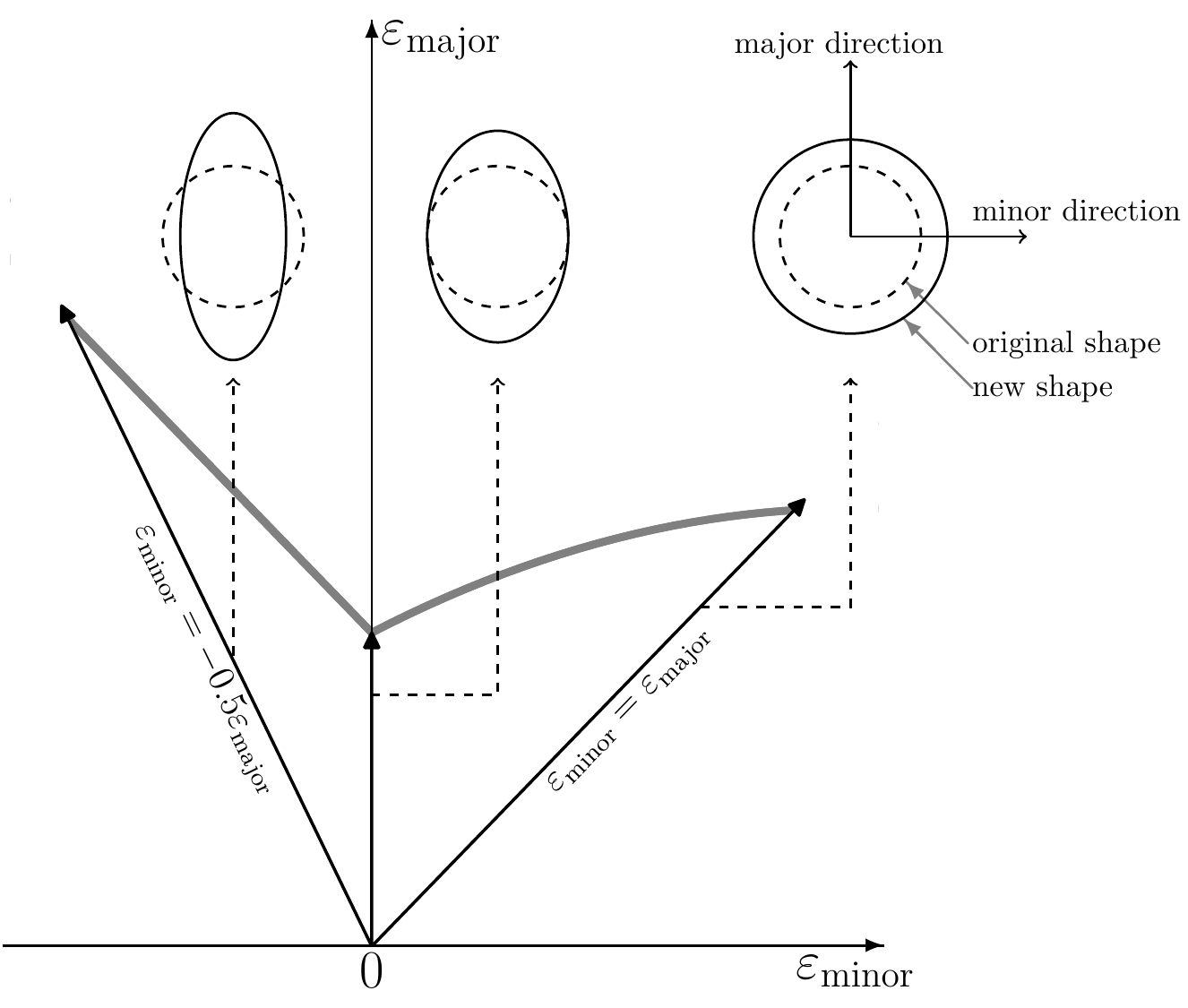}
\subcaption{}
\label{fig:intro_fld_2}
\end{subfigure}
\caption{(a) A schematic of a forming limit diagram (FLD). (b) Strain paths on a FLD.}
\label{fig:intro_fld}
\end{figure*}

In sheet metal forming, it is popular to use forming limit diagrams (FLDs) to guide forming operations.
A schematics of FLD is shown in \cref{fig:intro_fld_1} pioneered by Keeler \cite{Keeler1968} and Goodwin \cite{Goodwin1968}. An FLD basically maps ``safe'' and ``failure'' areas of the applied principle strains, \Emajor~ and \Eminor, along certain strain paths (see \cref{fig:intro_fld_2}). In \cref{fig:intro_fld_1}, the safe area is below the critical strain and the failure area is above it. The critical strain in \cref{fig:intro_fld_1} is not the strain for fracture, but the strain at which the localized necking occurs. The strain for fracture is usually above the localized necking, and approximately linearly depends on \Eminor~with a negative slope \cite{Hosford1999}. When the strain path is close to equi-biaxial tension condition, \ie\, \Emajor$=$\Eminor, fracture might occur without any localized necking event \cite{LeRoy1978}. 

An FLD is valid only when the strain path is linear. If the strain path is changed or non-linear, the FLD will shift \cite{Graf1993,vanHoutte2003,vanHoutte2007}. The strain path effect is pertinent when the shape of the formed component is complex or it undergoes multiple forming steps. Ignoring the strain path effect gives an inaccurate measure of the forming limits. 

There are two microstructural aspects regarding the strain path effect: the first one is the crystallographic texture evolution which changes the shape of the yield locus, and in turn it influences the final FLD \cite{Marciniak1978,Lian1989b}; the second aspect is the development of dislocation patterns during plastic deformation which cause cross effect or Bauschinger effect upon strain path change (see the seminal work by Teodosiu and co-workers \cite{Teodosiu1998,Teodosiu2002}). Physically based models incorporating those two microstructural aspects have been developed to model the strain path effect \cite{vanHoutte2003,vanHoutte2007,Hiwatashi1997,Hiwatashi1998}.

The strain path effect also gave rise to the development of stress-based FLD \cite{Stoughton2000,Stoughton2004,Wu2005,Yoshida2007b,Stoughton2012} in which the critical stress for localized necking is not sensitive to the strain path change. Despite the advantage of the stress based FLD over the conventional strain based FLD, accurate measurement of stress is rather challenging, whereas measuring strain can be very accurate.
Thus, the use of stress based FLD has not yet become a common engineering practice. 

\subsection{Theoretical models for forming limits}
Theoretically modeling FLD has been an active topic, and there are three popular families of theoretical models, \ie\ the maximum force criteria (MFC), the Marciniak-Kuczy\'{n}ski model (MK), and the bifurcation theory (BT). Comprehensive reviews can be found in \cite{Banabic2000,Banabic2010a,Abed-Meraim2014,Banabic2016}. Here, we give a brief sketch of them.
\subsubsection{Maximum force criteria (MFC)}
Among the maximum force criteria (MFC) family, the first model can be traced back to Consid\`{e}re \cite{Considere1885} who proposed the well-known Consid\`{e}re criterion, specifically for uniaxial tensile deformation
, \ie\,
\begin{subequations}
\begin{align}
\text{d}\vctr{f}_{1}=&0\\
\intertext{or}
\totalder{\sigma_{1}}{\varepsilon_{1}}=&\sigma_{1}
\label{eq:considere}
\end{align}
\end{subequations}
where $\vctr{f}_{1}$ is the force in the tensile direction; and $\sigma_{1}$ and $\varepsilon_{1}$ are the corresponding respective Cauchy stress and strain.

This 1D formulation by Consid\`{e}re was later adopted by Swift \cite{Swift1952}, and extended to 2D case, \ie\ biaxial loading, which is more relevant to sheet metal forming. 
Swift's hypothesis reads:
\begin{subequations}
\begin{align}
\text{d}\vctr{f}_{1}=0&\text{~~~~and~~~~}\text{d}\vctr{f}_{2}=0 \\
\intertext{or}
\totalder{\sigma_{1}}{\varepsilon_{1}}=\sigma_{1}&\text{~~~~and~~~~}\totalder{\sigma_{2}}{\varepsilon_{2}}=\sigma_{2}
\end{align}
\end{subequations}
Swift's hypothesis is, however, rarely supported by experimental observations \cite{Abed-Meraim2014}. Both Consid\`{e}re and Swift's criteria consider the case of diffuse necking. 

Also based on Consid\`{e}re's idea, Hill proposed a criteria for localized necking \cite{Hill1952}. In Hill's criterion, two conditions should be fulfilled when a localized necking occurs. The first condition is that the direction of the band where localized necking occurs is the same to the direction where the elongation is zero; the second condition is that in the normal direction of the localized band the force becomes extremum. These two conditions read:
\begin{subequations}
\begin{align}
\text{d}\varepsilon_{\text{tt}}=&0\\
\intertext{and}
\text{d}\vctr{f}_{\text{n}}=&0\text{~~~or~~~}\totalder{\sigma_{\text{n}}}{\varepsilon_{\text{n}}}=\sigma_{\text{n}}
\label{eq:hill52}
\end{align}
\end{subequations}
where n and t are the normal direction of the band and the direction along the band, respectively.

The MFC were further brought forward by Hora and co-workers \cite{Hora1996,Hora2013} based on the fact that localized necking takes place in plane strain condition. At the onset of localized necking, the deformation path represented by the incremental strain ratio should become zero, namely:
\begin{align}
\beta=\totalder{\varepsilon_{2}}{\varepsilon_{1}}=&0
\label{eq:hora1996_1}
\end{align}
Thus the criterion is 
\begin{align}
\partialder{\sigma_{1}}{\varepsilon_{1}}+\partialder{\sigma_{1}}{\beta}\partialder{\beta}{\varepsilon_{1}}=&\sigma_{1}
\label{eq:hora1996_2}
\end{align}
Though there are the numerical issues with Hora's criterion when straight line segments exist on the yield surface \cite{Aretz2004}, they are not directly caused by Hora's criterion itself, but by the yield surface description \cite{Manopulo2015}. Recently, Hora's criterion has been reformulated to be more flexible with the yield functions, also extended to include strain rate sensitivity and anisotropic hardening effects \cite{Manopulo2015}.
\subsubsection{Marciniak-Kuczy\'{n}ski model (MK-model)}
The Marciniak-Kuczy\'{n}ski model (MK-model) \cite{Marciniak1967} is physically intuitive. It assumes a pre-existing geometrical imperfection in the sheet metal, and it gradually evolves to localized necking as the deformation progresses. The imperfection is usually modeled as a thin band, and geometrically characterized by (i) its orientation relative to the principle directions of the sheet metal and (ii) its reduced thickness \cite{Banabic2010}. 

The basic assumption in the MK-model implicitly separates the sheet metal into two regions, \ie\, the region inside the band and the region outside the band. During deformation, two conditions are to be fulfilled \cite{Banabic2016a}: (i) the continuity of the strain rate along the band; (ii) the mechanical equilibrium on the interface between the above mentioned two regions. 

Localized necking occurs when a criterion (usually empirical) is met by comparing the kinematic evolutions in those two regions. This criterion can be the ratio of the strain rates in the major principal direction or the thickness direction, or the ratio of the equivalent strain rates \cite{Abed-Meraim2014}, and it can also be the ratio of the principal strains \cite{Banabic2010}. 

Since those criteria which detect the occurrence of localized necking only involve the kinematic evolution, the constitutive law is independent of the localized necking. Hence, the MK-model can be coupled with any constitutive laws, which lets the MK-model gain its flexibility in terms of modeling.

The postulate of a pre-existing geometrical imperfection in the MK-model is often criticized for being arbitrary, and the prediction is very sensitive to the two assumed geometrical characteristics of the pre-existing geometrical imperfection. The numerical value of the band orientation can be chosen in such a way that it corresponds to the minimum forming limit for a given strain path, and those minimum corresponding forming limits have been shown to be in good agreement with experiments (see for example in Ref. \cite{Signorelli2009a}). The value of the reduced thickness, on the other hand, can never be chosen in such a systematical way, and usually it is obtained by fitting the experiments. Even by fitting the same experiments, the obtained reduced thickness depends on the employed constitutive law \cite{Signorelli2009a}, indicating the reduced thickness is hardly a physical quantity.
\subsubsection{Bifurcation theory (BT)}
The bifurcation theory (BT) is theoretically and mathematically rigorous and sound \cite{Rudnicki1975,Stoeren1975,Rice1976,Rice1980,Forest2004}. Its essence is to inspect the stiffness matrix. It has certain similarities to the MK-model that a localized necking occurs also in a thin band rendering two regions inside and outside the band. The occurrence of the band and its geometrical characteristics are the predictions of BT, whereas in MK-model, the geometrical characters are assumed as model parameters. 

Due to the similarities between BT and the MK-model, BT has shown to be the upper limit of MK-model, \ie\, when the thicknesses inside and outside the band are the same \cite{Abed-Meraim2014}. This actually implies that the prediction by BT is likely to be overestimated, if assuming a reduced thickness would allow a good agreement with experiments. 

In principle, BT can be coupled with any constitutive laws. Coupling with rate dependent constitutive laws would, however, gives an unrealistic prediction, \eg\, the stress level when the localized necking occurs could be unrealistically high \cite{Yoshida2012}. An interested reader can also refer to All the instances are shown in the Supplementary Materials for the mathematical origin that coupling BT with a rate dependent constitutive law would give an unrealistic prediction.
\subsection{Previous studies of using crystal plasticity for FLD prediction}
According to the methods of realizing polycrystal simulations, the previous studies can be further cast into two groups: mean field and full field models. 
\subsubsection{Mean field models}
In the group of mean field models, the classical Taylor model \cite{Zhou1995,Wu1997,Savoie1998,Wu1998,Wu2004,McGinty2004,Inal2005,Yoshida2007,Wang2011,Yoshida2012} and the viscoplastic self-consistent model \cite{Signorelli2009,Signorelli2009a,Neil2009,Serenelli2010,Yang2010,Serenelli2011,Signorelli2012,Kim2013,Bertinetti2014,Franz2013,Schwindt2015,Akpama2017,Gupta2018} are often applied. Essentially, a crystal plasticity formulation together with a numerical homogenization scheme provides the stress-strain relation of a polycrystalline material, \ie\, a constitutive description.  As mentioned above, both the MK-model and BT are very flexible in terms of the constitutive law. Hence, these mean field polycrystal models are usually coupled with the MK-model \cite{Zhou1995,Wu1997,Savoie1998,Wu1998,Wu2004,McGinty2004,Inal2005,Yoshida2007,Wang2011,Yoshida2012,Signorelli2009,Signorelli2009a,Neil2009,Yang2010,Serenelli2010,Serenelli2011,Signorelli2012,Franz2013,Kim2013,Bertinetti2014,Schwindt2015,Akpama2017,Gupta2018} or BT \cite{Yoshida2012,Franz2013,Akpama2017}.
\subsubsection{Full field models}
As for the group of the full field model, it has been so far exclusively based on crystal plasticity finite element method in 2D \cite{Wu2004a,Viatkina2005,Inal2008} and 3D \cite{Lloyd2017,Kim2017,Mohammed2018}, and applying full field crystal plasticity models for FLD prediction deserves some discussion. 

It should be firstly pointed out that the primary purpose of applying any full field models is to reveal the physical process at a lower length scale. In the case simulating the localized deformation, the element size should be in general smaller than the size of the localized deformation. On the other hand, MFC and BT are both at the length scale of the components. It is in principle feasible to apply the MFC and BT in conjunction with the full field model, but it may lose all the essence of applying the full field models in the first place. 

For sheet metal forming simulations, it is tempting to use 2D models, but there are some issues. The first issue is the 2D representation of a polycrystalline material. In some previous studies, \eg\ \cite{Wu2004a,Inal2008}, only one orientation is assigned to each integration point, meaning through the sheet thickness, there is only one grain. This kind of 2D representation of a polycrystalline sheet is not only unrealistic, but also the observed localization is exaggerated \cite{Zeghadi2007a,Zeghadi2007b}. A more realistic 2D realization is to assign a grain aggregate to each integration point as in \cite{Viatkina2005}. 

The second issue is the fulfillment of the plane stress condition for sheet metal forming. It should be firstly pointed out that the single crystalline crystal plasticity formulation is not compatible with plane stress condition (see Section 8.3 in Ref. \cite{Roters2010} for more detailed clarification). While, in the studies of \cite{Wu2004a,Inal2008}, only the out-of-plane normal stress is restricted to be zero, and there is no mentioning of any additional constrains on the shear stress components and the shear strains. Thus, it is not clear whether the plane stress condition is strictly fulfilled in these previous studies.

To avoid the above issues in 2D models, it is more appropriate to use 3D models with proper boundary conditions to fulfill the macroscopic plane stress condition.
In the previous studies with 3D models \cite{Lloyd2017,Kim2017,Mohammed2018}, full field crystal plasticity 3D models are all coupled with the MK-model, \ie\, introducing a geometrical imperfection in a finite element method mesh. Direct coupling the MK-model with finite element method is rather straightforward, but this kind of simulations requires considerable computational resources. For example, in Ref. \cite{Lloyd2017}, a few hundred simulations were performed varying the orientation angle and depth of the initial geometrical imperfection. Thus, simplifications are sometimes made to reduce the computational effort \cite{Kim2017,Mohammed2018}.
\subsection{Motivation and scope of this study}
\label{sec:intro_motivation}
The idea of introducing a geometrical imperfection in those above mentioned previous studies with 3D models \cite{Lloyd2017,Kim2017,Mohammed2018} is to trigger a localization event. On the other hand, a localization event can be also triggered by microstructural heterogeneity, and polycrystalline materials are intrinsically heterogeneous. Hence, without a geometrical imperfection, a localized deformation event should naturally take place in a full field crystal plasticity simulation of a polycrystalline material, as already observed in many previous studies, \eg\, in 2D plane strain simulations \cite{Inal2002,Inal2002a,Inal2002b,Neale2003,Wu2007,Yoshida2014,Wu2017} or 3D uniaxial tension simulations \cite{Kim2015,Hu2016}.

To simulate localized deformation, it is not necessary to introduce an initial geometrical imperfection to prompt localized deformation when using full field crystal plasticity models. More importantly, the geometrical characteristics of the initial imperfection assumed in the MK-model are not measurable quantities, whereas microstructural can be always measured and characterized by modern analytic techniques.

Motivated by ``virtual material testing laboratory'' \cite{Raabe2002,Kraska2009,Helm2011,Zhang2015,Butz2016,Zhang2016} and ``integrated computational materials engineering'' (ICME) \cite{Konter2016} where physically and mechanistically based models are applied to minimize parameter fitting from experiments,
the aim of this study is at assessing the feasibility of using full field polycrystalline crystal plasticity models for the prediction of FLD. The localization event is to be triggered by microstructural heterogeneity, instead of assuming an initial geometrical imperfection.

We confine our focus here on the right hand side of the FLD of face centered cubic (fcc) metals. The rest of the text is organized as the follows: in \cref{sec:method}, we describe the employed crystal plasticity finite element method and the simulation setup. 
In \cref{sec:yoshida2007}, we compare the simulated FLDs by using 3D full field crystal plasticity model with a previous simulation study by Yoshida \textit{et al.} \cite{Yoshida2007} where the MK-model in conjunction with Taylor model was employed. In \cref{sec:wu1998}, we compare our simulations with a measured FLD from Ref. \cite{Wu1998}. In \cref{sec:limitation}, some limitations of this study are discussed, and finally this study is summarized in \cref{sec:summary}.
\section{Methodology}
\label{sec:method}
\subsection{Crystal plasticity}
\label{sec:cp}
In this study, crystal plasticity finite element method (CP-FEM) is applied. CP is to incorporate the microscopic plastic deformation mechanisms into continuum mechanics, and CP-FEM is to use FEM as a numerical solver to solve mechanical equilibrium equations for the solids whose constitutive behavior is formulated by CP. The subject of CP-FEM is very well documented in \cite{Roters2010,Roters2010a}, here a brief description is provided.
\subsubsection{Kinematics}
\label{sec:cp_kinematics}
In the present crystal plasticity formulation, the \term{deformation gradient}, $\F$, is decomposed into a plastic part and a elastic part,
\begin{align}
\F=\Fe\Fp
\label{eq:F}
\end{align}
where $\Fp$ is the \term{plastic deformation gradient}, and $\Fe$ is the \term{elastic deformation gradient}. The above decomposition is to firstly deform the solid to an intermediate configuration transformed by $\Fp$ from the reference configuration. During this transformation, the solid is deformed plastically, thus being stress free. For convenience, $\Fp$ contains a rotation part which is equivalent to the rotation matrix of the crystallographic orientation. Consequently, the intermediate configuration corresponds to the crystal frame. 

The transformation from the intermediate configuration to the current configuration is carried by $\Fe$. It contains the elastic stretching and a rotation part. The elastic stretching part is very close to a unity tensor for metals. The rotation part is to transform from the crystal frame to the sample frame. Thus, the rotation part in $\Fe$ is the inverse of the rotation matrix of the crystallographic orientation.

With \cref{eq:F}, it is easy to show that 
\begin{align}
\velgrad=\Le+\Fe\Lp\inverse{\Fe}
\label{eq:L}
\end{align}
where $\velgrad$ is the \term{velocity gradient} corresponding to $\F$, and $\Le$ and $\Lp$ are the respective velocity gradients for $\Fe$ and $\Fp$. The \term{plastic velocity gradient}, $\Lp$, is a function of the stress measure at the intermediate configuration and the current microstructure,
\begin{align}
\Lp=f\left(\sPK,S_{1},S_{2},\cdots\right)
\label{eq:Lp}
\end{align}
where $\sPK$ is the \term{second Piola-Kirchhoff stress}, and $S_{i}$ are the state variables describing the microstructure. \cref{eq:Lp} is essentially a constitutive law, and a specific form of it employed in this study will be given in the next section. 
\subsubsection{Constitutive law}
\label{sec:cp_constitutive}
In this study, we confine ourselves with the fcc crystal structure and only consider the dislocation slip as the plastic deformation mechanism. In this case, the specific form of \cref{eq:Lp} is
\begin{align}
\label{eq:Lp_dot_gamma}
\Lp=\sum_{\alpha=1}^{12}\dot{\gamma}^{\alpha}\left(\vctr{m}^{\alpha}\otimes\vctr{n}^{\alpha}\right)
\end{align}
where $\dot{\gamma}^{\alpha}$ is the shear rate on the slip system $\alpha$, and its shear direction is $\vctr{m}^{\alpha}$, shear plane normal is $\vctr{n}^{\alpha}$. The dyadic product, $\vctr{m}^{\alpha}\otimes\vctr{n}^{\alpha}$, is termed as \term{Schmid tensor}.

As proposed in Ref. \cite{Peirce1983}, $\dot{\gamma}^{\alpha}$ depends on the resolved shear stress, $\tau^{\alpha}$ on the shear plane, and the shear resistance, $\tau_{c}^{\alpha}$, of this slip system,
\begin{align}
\label{eq: shear rate pheno}
\dot\gamma^\alpha &= \dot\gamma_0\abs{\frac{\tau^\alpha}{\tau_{c}^\alpha}}^{\frac{1}{m}} \sign{\tau^\alpha}\
\end{align}
where $\dot\gamma_0$ and $a$ are material parameters. 

$\tau^\alpha$ is obtained by projecting $\sPK$ on the corresponding Schmid tensor,
\begin{align}
\tau^\alpha = \sPK \doublecon \left(\vctr{m}^{\alpha}\otimes\vctr{n}^{\alpha}\right)
\end{align}
in which $\sPK$ is related to the elastic deformation gradient, $\Fe$, by
\begin{align}
\label{eq: sPK}
\sPK = \tnsrfour{C}\doublecon(\transpose{\Fe}\Fe-\eyetwo)/2
\end{align}
where $\tnsrfour{C}$ is the \term{elastic stiffness tensor}.

In this constitutive description, the microstructure is solely characterized by the slip resistance on the slip system, $\tau_{c}^{\alpha}$, and it saturates towards $\tau_\infty^\alpha$ due to slip activity on systems $\beta$,
\begin{align}
\label{eq: hardening pheno}
\dot \tau_{c}^\alpha &= h_0\left(1-\frac{\tau_{c}^\alpha}{\tau_\infty^\alpha}\right)^w h_{\alpha\beta}\,\left|\dot\gamma^\beta\right|
\end{align}
where $h_{\alpha\beta}$ is a 12$\times$12 matrix in which 1.0 is for the case when $\alpha$ and $\beta$ are coplanar slip systems, and 1.4 otherwise. $h_{0}$ and $w$ are material parameters.
\subsection{Crystal plasticity finite element method (CP-FEM)}
The CP-FEM simulations are realized by incorporating a CP simulation package, \ie\, DAMASK (D{\"u}sseldorf Advance MAterials Simulation Kit) \cite{damask_website,Roters2012,Roters2019}, with a commercial FEM simulation software, LS-DYNA \cite{lsdyna}. The interface between DAMASK and LS-Dyna is created via the user defined material property subroutine in LS-Dyna. The algorithm is briefly described as the follows. 

At the the beginning, LS-DYNA passes a new deformation gradient, $\F$ and a new time step, $\Delta t$ into the user defined material subroutine. By using a guessed $\Fp$ and \cref{eq:F}, the elastic deformation gradient, $\Fe$ is obtained, which in turn is used to evaluate the second Piola-Kirchhoff stress, $\sPK$ (\cref{eq: sPK}). Then, a $\Lp$ is obtained by the using the constitutive law described in the preceding section together with the newly evaluated $\sPK$. With the knowledge of $\Lp$, the guessed $\Fp$ is updated by using its rate, \ie\, $\dot{\F}_{\text p}=\Lp\Fp$ and the given $\Delta t$. The updated $\Fp$ is then used to update $\sPK$, then update $\Lp$, which goes on to make a new update on $\Fp$. This loop continues until the variables, $\Fp$, $\sPK$, and $\Lp$, are consistent with each other for the given $\F$ and the given time step $\Delta t$. 

\begin{figure*}[h!]
\centering
\includegraphics[scale=0.8]{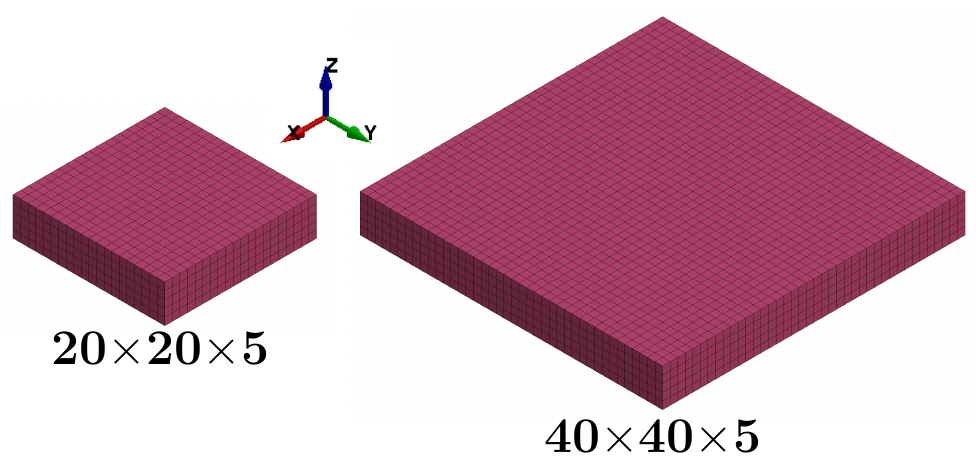}
\caption{Two meshes used for the simulations. 20$\times$20$\times$5 means there are 20 elements along the width, 20 along the breadth, and 5 in thickness.}
\label{fig:mesh}
\end{figure*}
\begin{figure*}
\centering
\begin{subfigure}[b]{0.9\linewidth}
    \centering
    \includegraphics[scale=0.25]{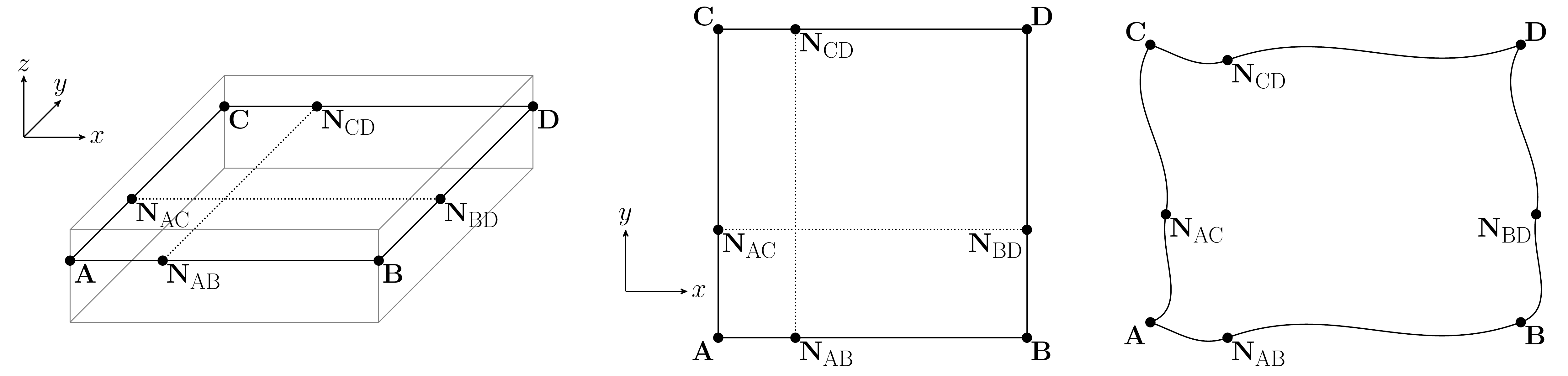}
    \subcaption{Periodic boundary condition on $x-y$ plane}
    \label{fig:pbc_xy}
\end{subfigure}%
\\
\begin{subfigure}[b]{0.9\linewidth}
    \centering
    \includegraphics[scale=0.25]{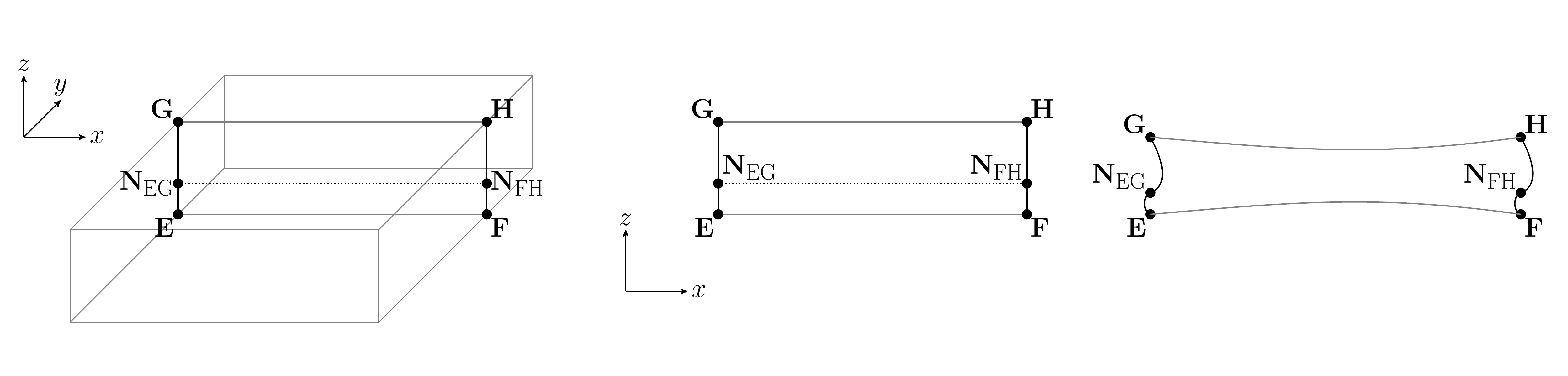}
    \subcaption{Periodic boundary condition on $z-x$ plane}
    \label{fig:pbc_zx}
\end{subfigure}
\begin{subfigure}[b]{0.9\linewidth}
    \centering
    \includegraphics[scale=0.25]{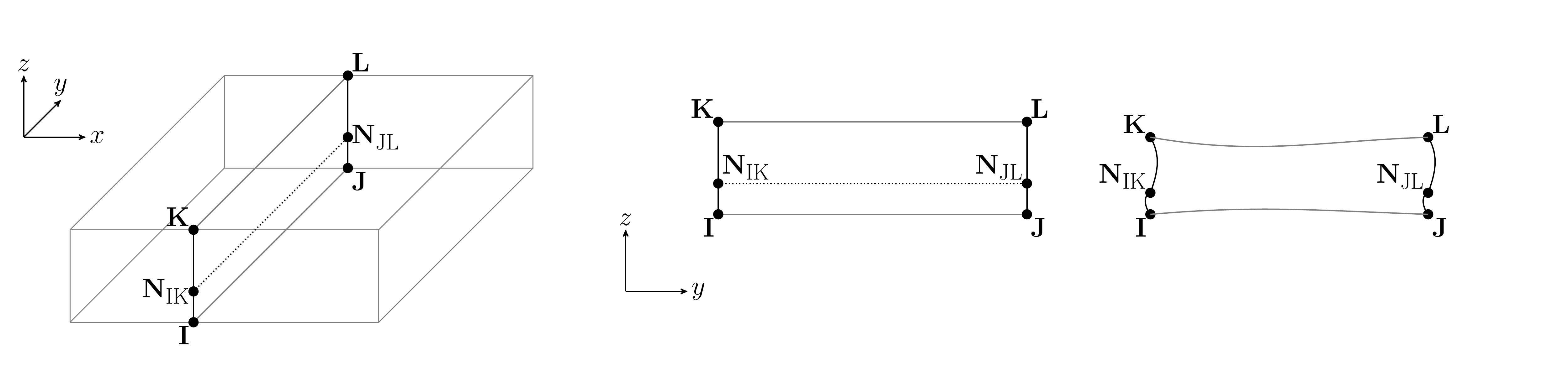}
    \subcaption{Periodic boundary condition on $y-z$ plane}
    \label{fig:pbc_yz}
\end{subfigure}
\begin{subfigure}[b]{0.9\linewidth}
    \centering
    \includegraphics[scale=0.25]{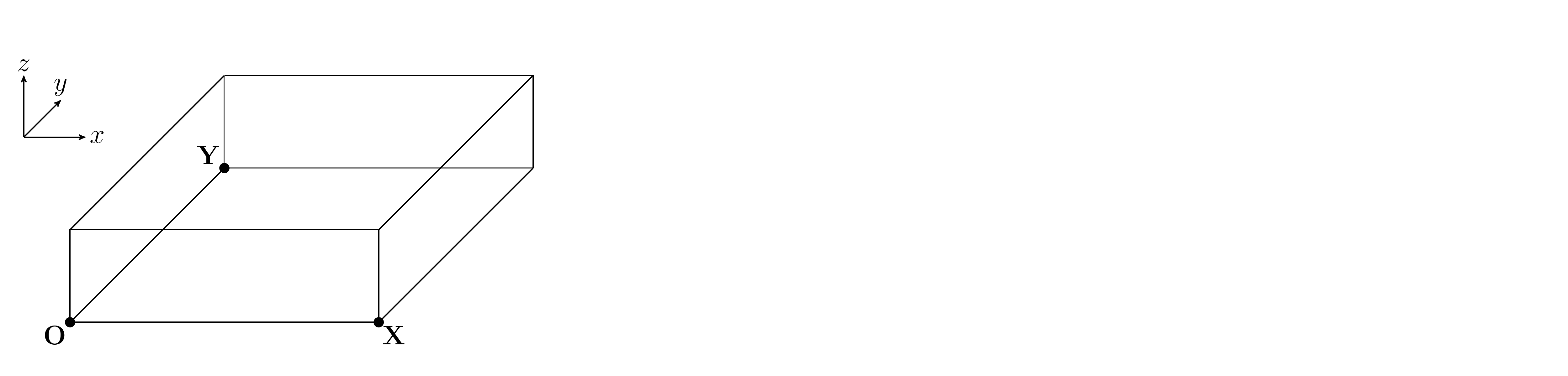}
    \subcaption{Locations of control nodes}
    \label{fig:bc}
\end{subfigure}
\caption{Periodic boundary condition applied on (a) $x-y$ (see \cref{eq:pbc_xy}), (b) $z-x$ (see \cref{eq:pbc_zx}), and (c) $y-z$ (see \cref{eq:pbc_yz}) planes, and (d) three control nodes.}
\label{fig:pbc_bc}
\end{figure*}

Besides the above consistent loop, another consistent loop is also performed during the process of obtaining $\Lp$ by using the given $\sPK$. With the specific constitutive law in the preceding section, $\dot\gamma^\alpha$ and $\dot\tau_{c}^\alpha$ are to be consistent with each other at the end of this consistent loop. In general, the obtained $\Lp$ is to be consistent with the state variable $S_{i}$ in \cref{eq:Lp} for any given constitutive laws at a given $\sPK$.

At the end of these consistent loops, the newly updated $\sPK$ and $\Fp$ are used to obtain the \term{first Piola-Kirchhoff stress}, $\fPK$, by
\begin{align}
\fPK=\F\inverse{\Fp}\sPK\invtranspose{\Fp}
\label{eq:S2P}
\end{align}
One may notice that the above expression is different from the usual conversion between $\fPK$ and $\sPK$, which is simply $\fPK=\F\sPK$. As described in \cref{sec:cp_kinematics}, $\sPK$ is the stress measure in the intermediate configuration transformed from the reference configuration by $\Fp$. Thus, $\sPK$ should be firstly transformed back from the intermediate configuration to the reference configuration.

Afterwards, the \term{Cauchy stress}, $\tnsr{\sigma}$, can be easily obtained by the usual conversion between $\fPK$ and $\tnsr{\sigma}$.
\begin{align}
\tnsr{\sigma}=\frac{1}{\text{det}\left(\F\right)}\fPK\transpose{\F}
\label{eq:P2sigma}
\end{align}
Then $\tnsr{\sigma}$ is returned to LS-DYNA. 

It should be pointed out that $\tnsr{\sigma}$ can be directly obtained by pushing-forward $\sPK$ with $\Fe$, \ie\, $\tnsr{\sigma}=\frac{1}{\text{det}\left(\F\right)}\Fe\sPK\transpose{\Fe}$, and the two-step procedure (\cref{eq:S2P,eq:P2sigma}) described here is not necessary. The main reason for this two-step procedure is that within DAMASK the mechanical equilibrium equation is operated on $\fPK$ in the reference frame, and it is provided by DAMASK as an output variable. Thus, $\tnsr{\sigma}$ required by LS-DYNA is obtained here by using the $\fPK$ from DAMASK and the $\F$ from LS-DYNA.

\subsection{Simulation setup}
Sheet metals under biaxial tension is modeled. We use two meshes as shown in \cref{fig:mesh}. The dimensions of the meshes are 4 mm$\times$ 4 mm $\times$ 1 mm and 8 mm$\times$ 8 mm $\times$ 1 mm (width$\times$breadth$\times$thickness). The element size is 0.2 mm $\times$ 0.2 mm $\times$ 0.2 mm, which renders the total number of elements to be 20$\times$20$\times$5 = 2000, and 40$\times$40$\times$5 = 8000, respectively. Full integration element with 8 integration points is used. A single orientation is assigned to each integration point.

Periodic boundary condition (PBC) is applied to the above meshes to mimic an infinitive sheet metal under biaxial tension in $x$ and $y$ directions (see \cref{fig:mesh} for the coordinate system). PBC is applied to the faces normal to $x$ or $y$ axis. The PBC is described as follows \cite{Sluis2000,Li2004}:
\begin{itemize}
\item{PBC on $x$-$y$ plane:
in \cref{fig:pbc_xy}, the nodes {\bf A}, {\bf B}, {\bf D}, and {\bf C} are on an arbitrary plane whose normal is parallel to the $z$ axis. The node {\bf N$_{\text{AB}}$} is an arbitrary node on {\bf A}--{\bf B}, and its equivalent node on {\bf C}--{\bf D} is {\bf N$_{\text{CD}}$}. The node {\bf N$_{\text{AC}}$} is another arbitrary node on {\bf A}--{\bf C}, and its equivalent node on {\bf B}--{\bf D} is {\bf N$_{\text{BD}}$}. The displacements, \vctr{u}, of these nodes should fulfill the following relations:
\begin{subequations}
\begin{align}
\vctr{u}_{\text{N}_{\text{AB}}}-\vctr{u}_{\text{A}}&=\vctr{u}_{\text{N}_{\text{CD}}}-\vctr{u}_{\text{C}}\\
\vctr{u}_{\text{N}_{\text{AC}}}-\vctr{u}_{\text{A}}&=\vctr{u}_{\text{N}_{\text{BD}}}-\vctr{u}_{\text{B}}\\
\vctr{u}_{\text{B}}-\vctr{u}_{\text{A}}&=\vctr{u}_{\text{D}}-\vctr{u}_{\text{C}}
\end{align}
\label{eq:pbc_xy}
\end{subequations}
} 
\item{PBC on $z$-$x$ plane:
in \cref{fig:pbc_zx}, the nodes {\bf E}, {\bf F}, {\bf H}, and {\bf G} are on an arbitrary plane whose normal is parallel to the $y$ axis. The node {\bf N$_{\text{EG}}$} is an arbitrary node on {\bf E}--{\bf G}, and its equivalent node on {\bf H}--{\bf F} is {\bf N$_{\text{FH}}$}. The displacements, \vctr{u}, of these nodes should fulfill the following relations:
\begin{subequations}
\begin{align}
\vctr{u}_{\text{N}_{\text{EG}}}-\vctr{u}_{\text{E}}&=\vctr{u}_{\text{N}_{\text{FH}}}-\vctr{u}_{\text{H}}\\
\vctr{u}_{\text{G}}-\vctr{u}_{\text{E}}&=\vctr{u}_{\text{H}}-\vctr{u}_{\text{F}}
\end{align}
\label{eq:pbc_zx}
\end{subequations}
Since the simulation is meant for sheet metals, nodes on {\bf E}--{\bf F} are not equivalent to the nodes on {\bf G}--{\bf H}.
}
\item{PBC on $y$-$z$ plane:
in \cref{fig:pbc_yz}, the nodes are constrained by the following conditions,
\begin{subequations}
\begin{align}
\vctr{u}_{\text{N}_{\text{IK}}}-\vctr{u}_{\text{I}}&=\vctr{u}_{\text{N}_{\text{JL}}}-\vctr{u}_{\text{J}}\\
\vctr{u}_{\text{K}}-\vctr{u}_{\text{I}}&=\vctr{u}_{\text{L}}-\vctr{u}_{\text{J}}
\end{align}
\label{eq:pbc_yz}
\end{subequations}
}
\end{itemize}

\cref{fig:bc} shows three control nodes, {\bf O}, {\bf X}, and {\bf Y}. The translation and rotation of {\bf O} is fully constrained. Prescribed displacements are applied to the nodes {\bf X} and {\bf Y} along the respective $x$ and $y$ position directions. The prescribed displacements on {\bf X} and {\bf Y} follow the relation that
\begin{align}
u_{\text{Y}}=&L_{0}\left(\left(1+\frac{u_{\text{X}}}{L_{0}}\right)^{\rho}\right)-L_{0}
\end{align}
where $L_{0}$ is the initial width or breadth of sheet unit cell (see \cref{fig:mesh}). This leads to 
\begin{align}
\ln\left(\frac{u_{\text{Y}}}{L_{0}}+1\right)=&\ln\left(\frac{u_{\text{X}}}{L_{0}}+1\right)^{\rho}\nonumber\\
\varepsilon_{\text{minor}}=&\rho\cdot\varepsilon_{\text{major}}
\label{eq:rho}
\end{align}
where $\rho$ prescribes the strain path, \eg\, $\rho$ = 0.0 for plane strain tension, and $\rho$ = 1.0 for equal biaxial tension.
\section{Simulated texture dependence of FLD in comparison with a previous simulation study in Ref. \cite{Yoshida2007}}
\label{sec:yoshida2007}
In this section, we focus on the FLDs of some typical texture components in fcc metal, and compare the simulated FLDs with a previous simulation study by Yoshida \textit{et al}. \cite{Yoshida2007} in which FLD is modeled by using the MK-model in conjunction with Taylor model.
\subsection{Simulation input: material parameters and orientation distributions}
\begin{figure}
\centering
\begin{subfigure}[b]{1\linewidth}
    \centering
    \includegraphics[scale=0.2]{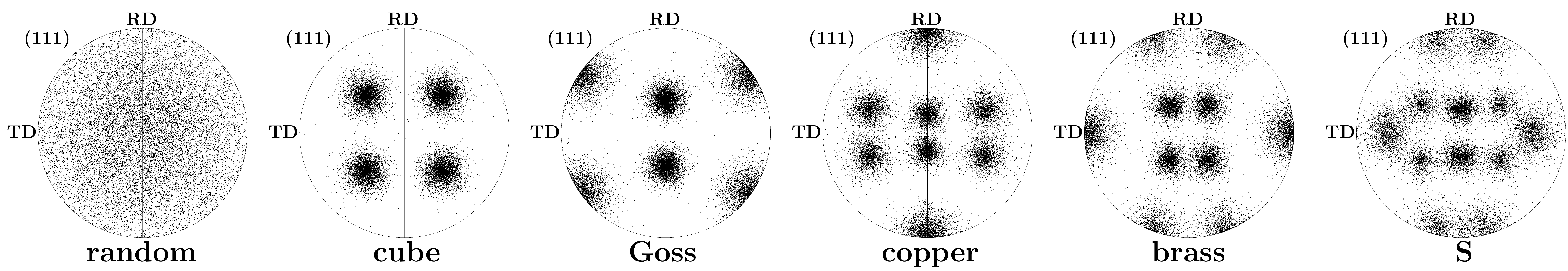}
    \subcaption{mesh 20$\times$20$\times$5}
    \label{fig:pf_20x20x5}
\end{subfigure}
\begin{subfigure}[b]{1\linewidth}
    \centering
    \includegraphics[scale=0.2]{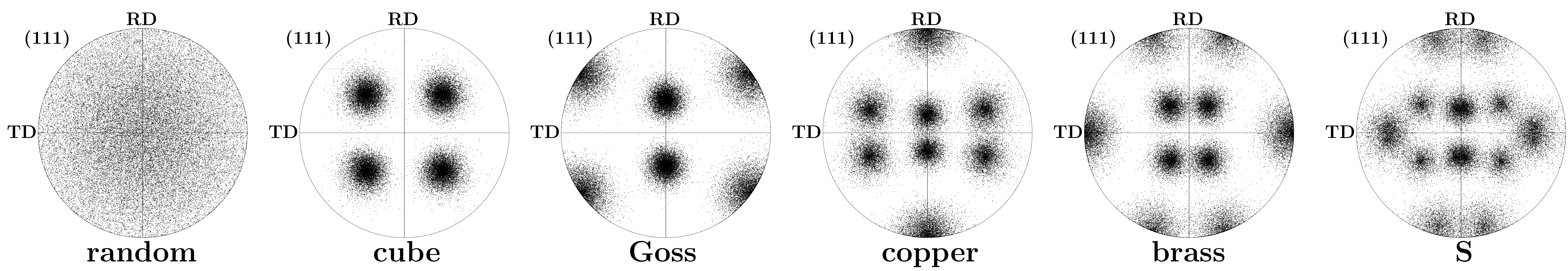}
    \subcaption{mesh 40$\times$40$\times$5}
    \label{fig:pf_40x40x5}
\end{subfigure}
\caption{Orientation distributions in the simulations for the meshes (a) 20$\times$20$\times$5 and (b) 40$\times$40$\times$5. The texture components are scattered by 15$^{\circ}$ around their ideal orientations. RD (rolling direction) coincides with $x$, and TD (transverse direction) coincides with $y$ in \cref{fig:mesh,fig:pbc_bc}. Note that 5 distribution instances for each orientation distribution and each mesh are used, and here we only show 1 instance for each mesh as examples. All the instances are shown in the Supplementary Materials. The orientation generation was realized by using the method described in \cite{Helming1998,Zhao2001,Raabe2004}. The pole figures are created by using ATEX \cite{jtex}}
\label{fig:pf}
\end{figure}

Six orientation distributions are considered: random, cube, Goss, copper, brass, and S. The latter five texture components are scattered around their ideal orientation by 15$^{\circ}$, as shown in \cref{fig:pf}. For each texture component, five instances are created for each mesh. In \cref{fig:pf}, only one instance is shown for each mesh, and in total 10 instances for each distribution. The pole figures of the other instances can be found in the Supplementary Materials. 
The materials parameters in \cref{eq: hardening pheno,eq: shear rate pheno,eq: sPK} are listed in \cref{tab:cp_parameters}. Those parameters are from \cite{Eisenlohr2013} for pure aluminium.

\begin{table}[h]
\centering
\caption{Material parameters in \cref{eq: hardening pheno,eq: shear rate pheno,eq: sPK}. These parameters are from \cite{Eisenlohr2013} for pure aluminium. Note $\alpha=1,\ldots,12$. $\tau^{\alpha}_{c0}$ is the initial shear resistance of the slip system $\alpha$. $^{\ast}$ if $\alpha$ and $\beta$ are coplanar slip systems. $^{\ast\ast}$ if $\alpha$ and $\beta$ are not coplanar slip systems.}
\label{tab:cp_parameters}
\begin{tabular}{lll|cr|crl}
\toprule
 $\tau^{\alpha}_{c0}$     & 31    & MPa & $\dot{\gamma_{0}}$ & 0.001 & $\tnsrfour{C}_{11}$ & 106.75 & GPa\\
 $\tau^{\alpha}_{\infty}$ & 64    & MPa & $m$                & 0.05    & $\tnsrfour{C}_{12}$ & 60.41  & GPa\\
 $w$                      & 2.25  &     &                    &       & $\tnsrfour{C}_{44}$ & 28.34  & GPa\\
 $h_{0}$ & 75 & MPa & & & & &\\
 $h_{\alpha\beta}$ & 1.0$^{\ast}$ & & & & & &\\
                   & 1.4$^{\ast\ast}$ & & & & & &\\
\bottomrule
\end{tabular}
\end{table}

\subsection{Identifying deformation localization and forming limit}
\begin{figure}[h!]
\begin{subfigure}[b]{0.5\linewidth}
    \centering
    \includegraphics[scale=0.3]{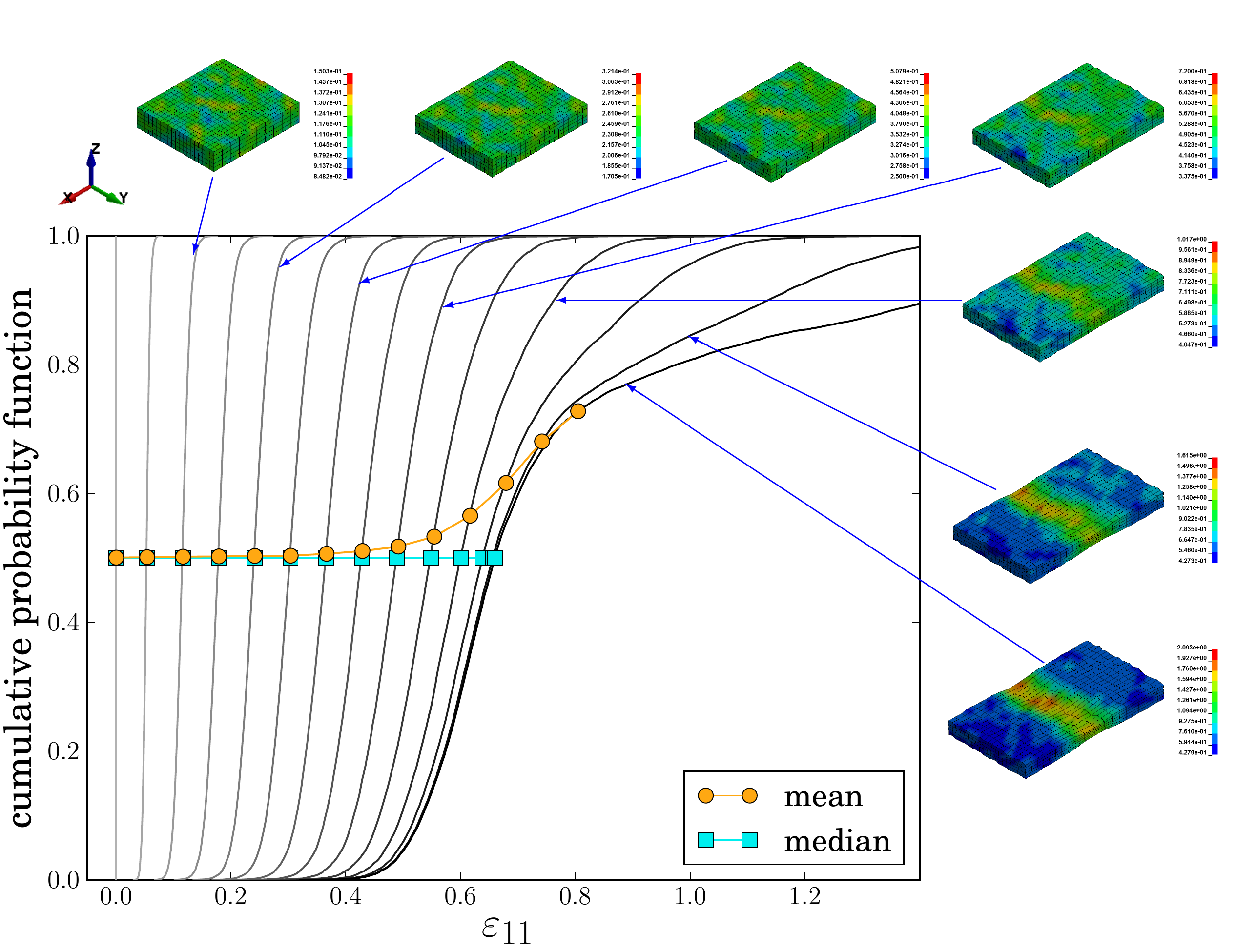}
    \subcaption{}
    \label{fig:r00_R1_random_e11}
\end{subfigure}
\begin{subfigure}[b]{0.5\linewidth}
    \centering
    \includegraphics[scale=0.3]{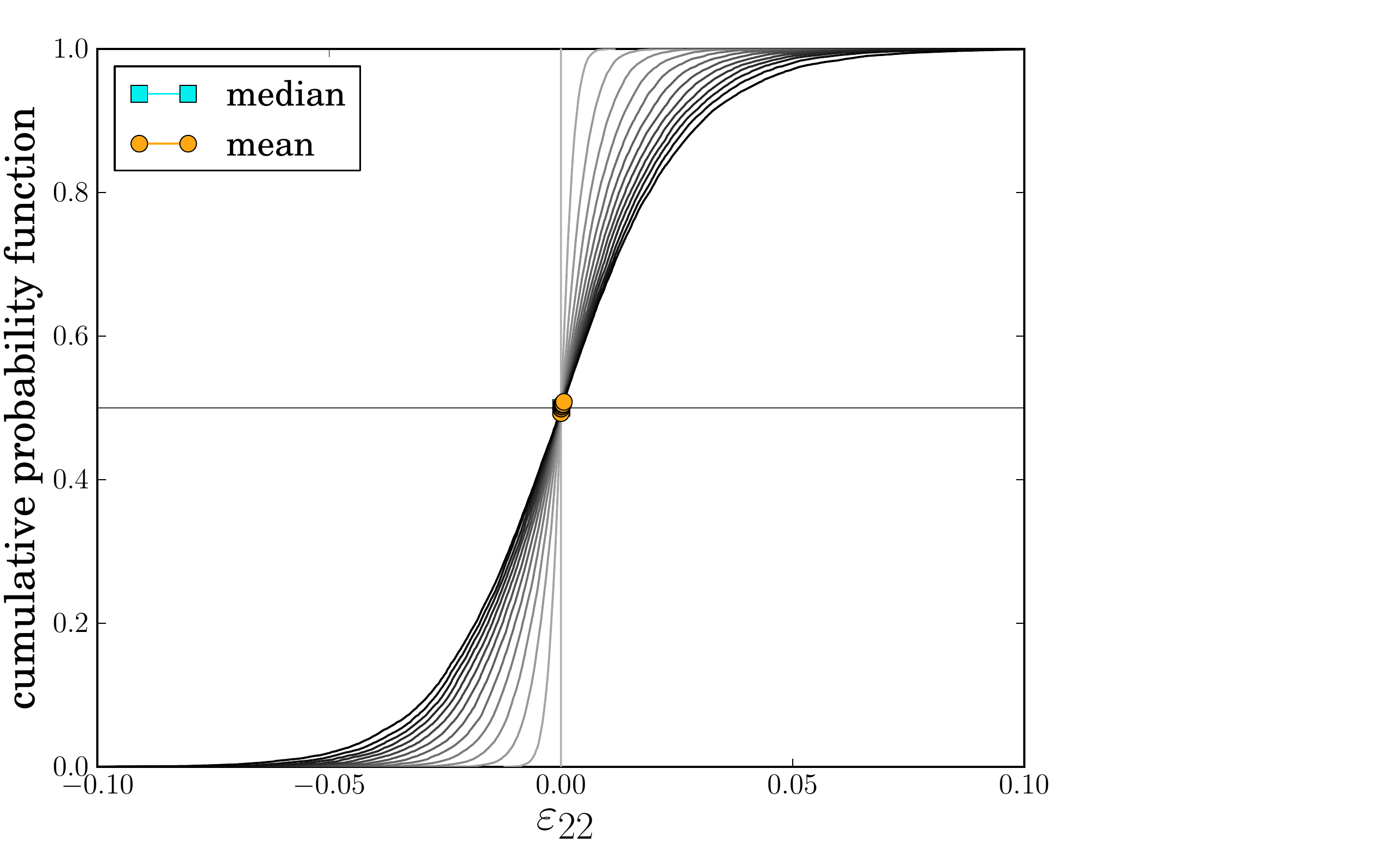}
    \subcaption{}
    \label{fig:r00_R1_random_e22}
\end{subfigure}
\caption{Evolution of the cumulative probability functions of (a) $\varepsilon_{11}$ and (b) $\varepsilon_{22}$ under plane strain tension, \ie\, $\rho=0.0$ (see \cref{eq:rho}). The orientation distribution is random and the mesh is 20$\times$20$\times$5.}
\label{fig:r00_R1_random_eii}
\end{figure}
\begin{figure}[h!]
\begin{subfigure}[b]{0.5\linewidth}
    \centering
    \includegraphics[scale=0.3]{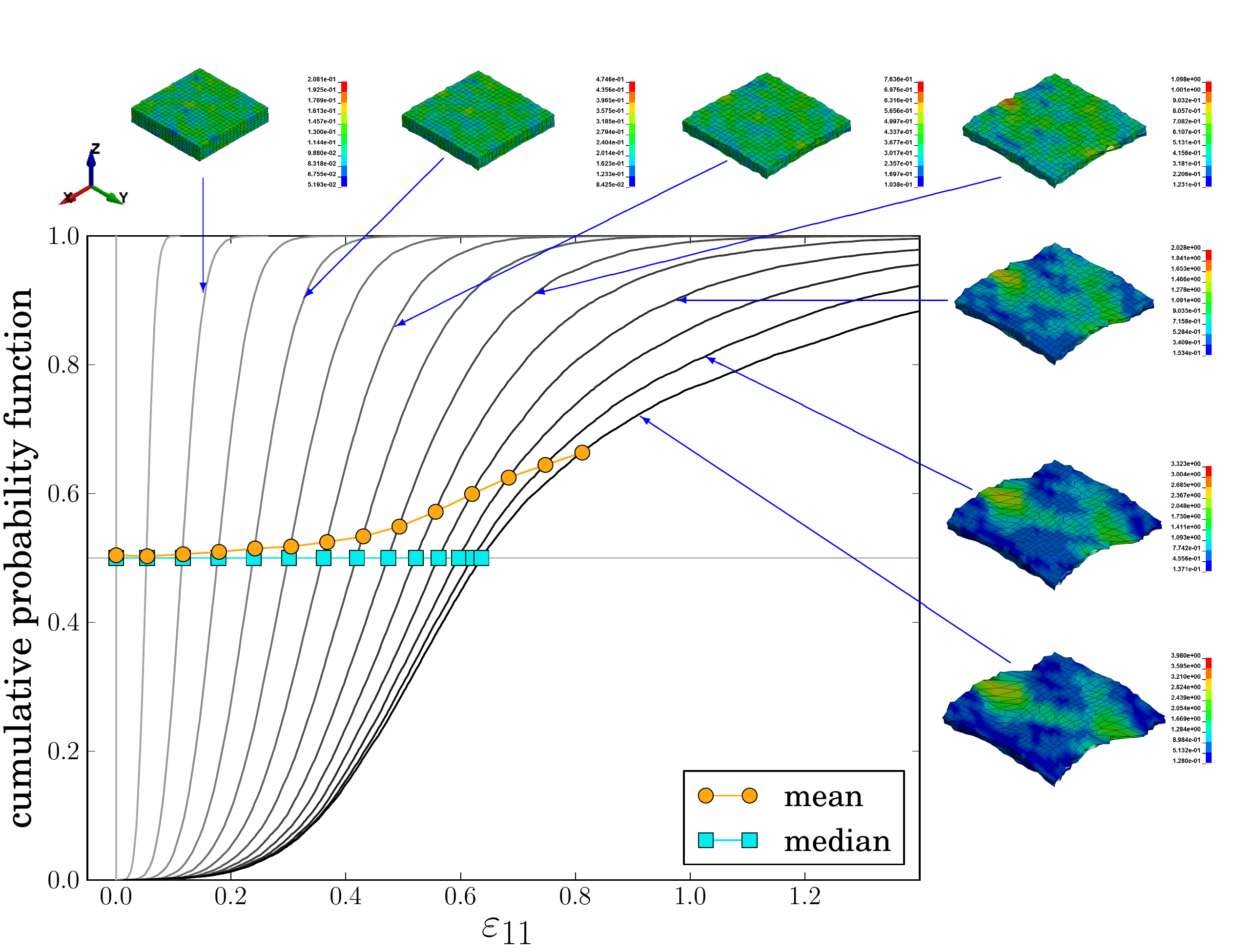}
    \subcaption{}
    \label{fig:r10_R1_random_e11}
\end{subfigure}
\begin{subfigure}[b]{0.5\linewidth}
    \centering
    \includegraphics[scale=0.3]{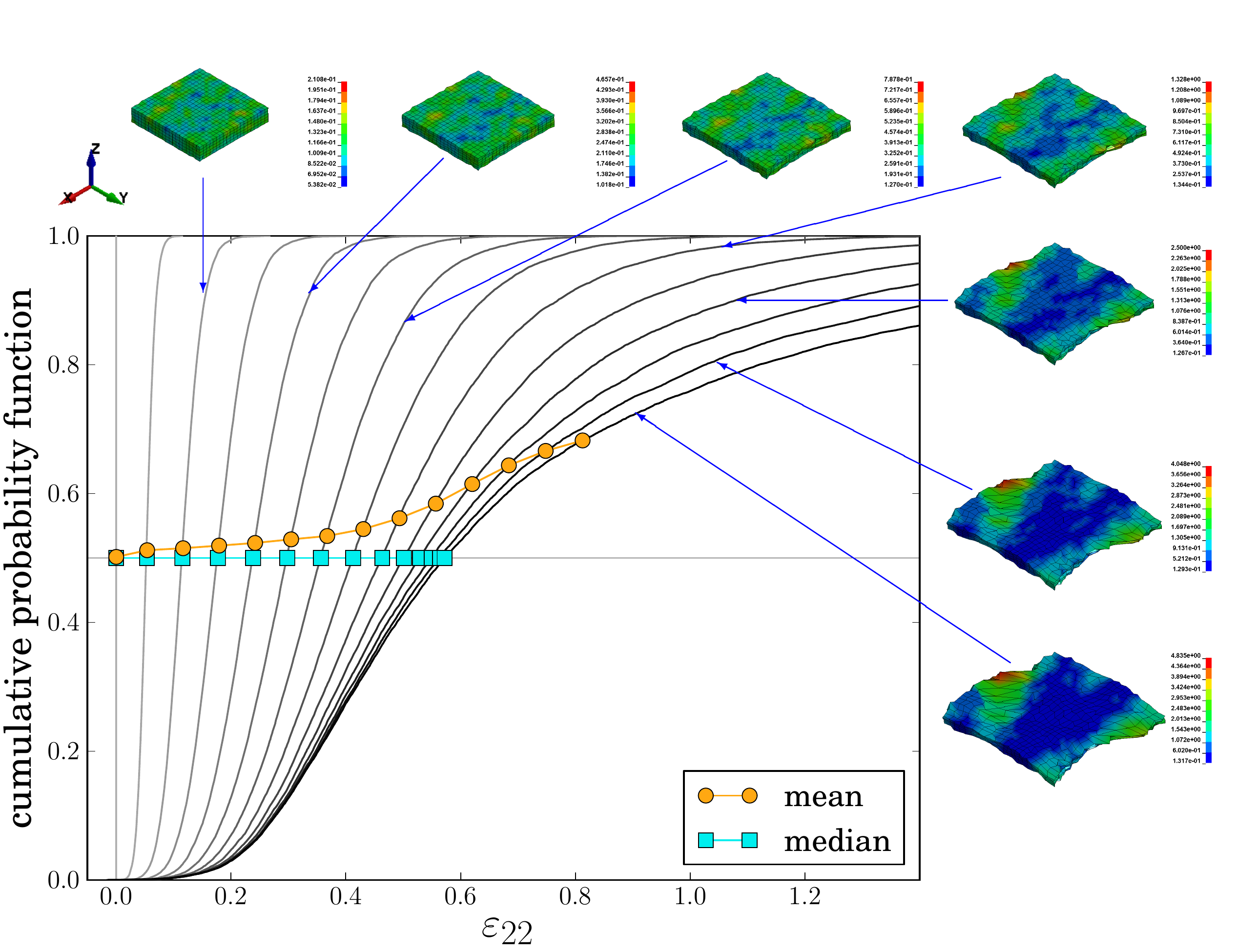}
    \subcaption{}
    \label{fig:r10_R1_random_e22}
\end{subfigure}
\caption{Evolution of the cumulative probability functions of (a) $\varepsilon_{11}$ and (b) $\varepsilon_{22}$ at equi-biaxial tension condition, \ie\ \Emajor=\Eminor, \ie\, $\rho=1.0$ (see \cref{eq:rho}). The orientation distribution is random and the mesh is 20$\times$20$\times$5.}
\label{fig:r10_R1_random_eii}
\end{figure}
Strictly speaking, there is no quantitative measure for localized deformation or necking. Thus, in this section, we discuss the measure employed in this study for quantifying localized deformation. Here, the localized deformation is quantified from a statistical point of view. To illustrate, we show two examples for one of the random texture distribution instances in \cref{fig:r00_R1_random_eii,fig:r10_R1_random_eii}, where the evolutions of the cumulative probability functions (CPFs) of $\varepsilon_{11}$ and $\varepsilon_{22}$ are shown. 

In \cref{fig:r00_R1_random_e11} when $\rho=0.0$ (plane strain tension and see \cref{eq:rho}), at early stage of deformation, the distribution of $\varepsilon_{11}$ approximately follows the normal distribution. This is because the mean value, $\varepsilon_{11}^{\text{mean}}$, is approximately the same to the median value, $\varepsilon_{11}^{\text{median}}$. As the deformation progresses, $\varepsilon_{11}^{\text{mean}}$ and $\varepsilon_{11}^{\text{median}}$ deviate from each other, indicating the distribution becomes more and more asymmetric. 

As $\varepsilon_{11}^{\text{mean}}$ and $\varepsilon_{11}^{\text{mean}}$ deviates apart, the population of $\varepsilon_{11}$ below $\varepsilon_{11}^{\text{median}}$ starts to evolve slowly, while the population above it advances quickly. Eventually, $\varepsilon_{11}^{\text{mean}}$ is mainly driven by the population above $\varepsilon_{11}^{\text{median}}$. These observations are consistent with the formation of a localized deformation in the direction of $\varepsilon_{11}$ (see the contour plots of $\varepsilon_{11}$ accompanying \cref{fig:r00_R1_random_e11}). 

\cref{fig:r00_R1_random_e22} shows that $\varepsilon_{22}^{\text{mean}}$ and $\varepsilon_{22}^{\text{median}}$ are always close to zero, which is consistent with the boundary condition of the plane strain tension ($\rho=0.0$). Though, as the deformation progresses, the distribution of $\varepsilon_{22}$ becomes wider, yet there is no indication of localized deformation. 

In \cref{fig:r10_R1_random_eii} for $\rho=1.0$ (equi-biaxial tension), the mean values of $\varepsilon_{11}$ and $\varepsilon_{22}$ both deviate from their respective median values. In comparison with the plane strain tension in \cref{fig:r00_R1_random_e11}, the deviations of the mean from the median values start at the earlier stage of the deformation. 

Based on the observations in \cref{fig:r00_R1_random_eii,fig:r10_R1_random_eii}, we use the ratios, $\varepsilon_{ii}^{\text{mean}}/\varepsilon_{ii}^{\text{median}}$ ($i$=1 or 2), as the measure for observing the evolution of localized deformation. As illustrated in \cref{fig:r00_R1_random_eii,fig:r10_R1_random_eii}, $\varepsilon_{ii}^{\text{mean}}/\varepsilon_{ii}^{\text{median}}$ = 1 indicates (approximately) homogenous deformation, and when it deviates from 1, the deformation becomes localized.

\begin{figure}[h!]
\begin{subfigure}[b]{1.0\linewidth}
    \centering
    \includegraphics[scale=0.55]{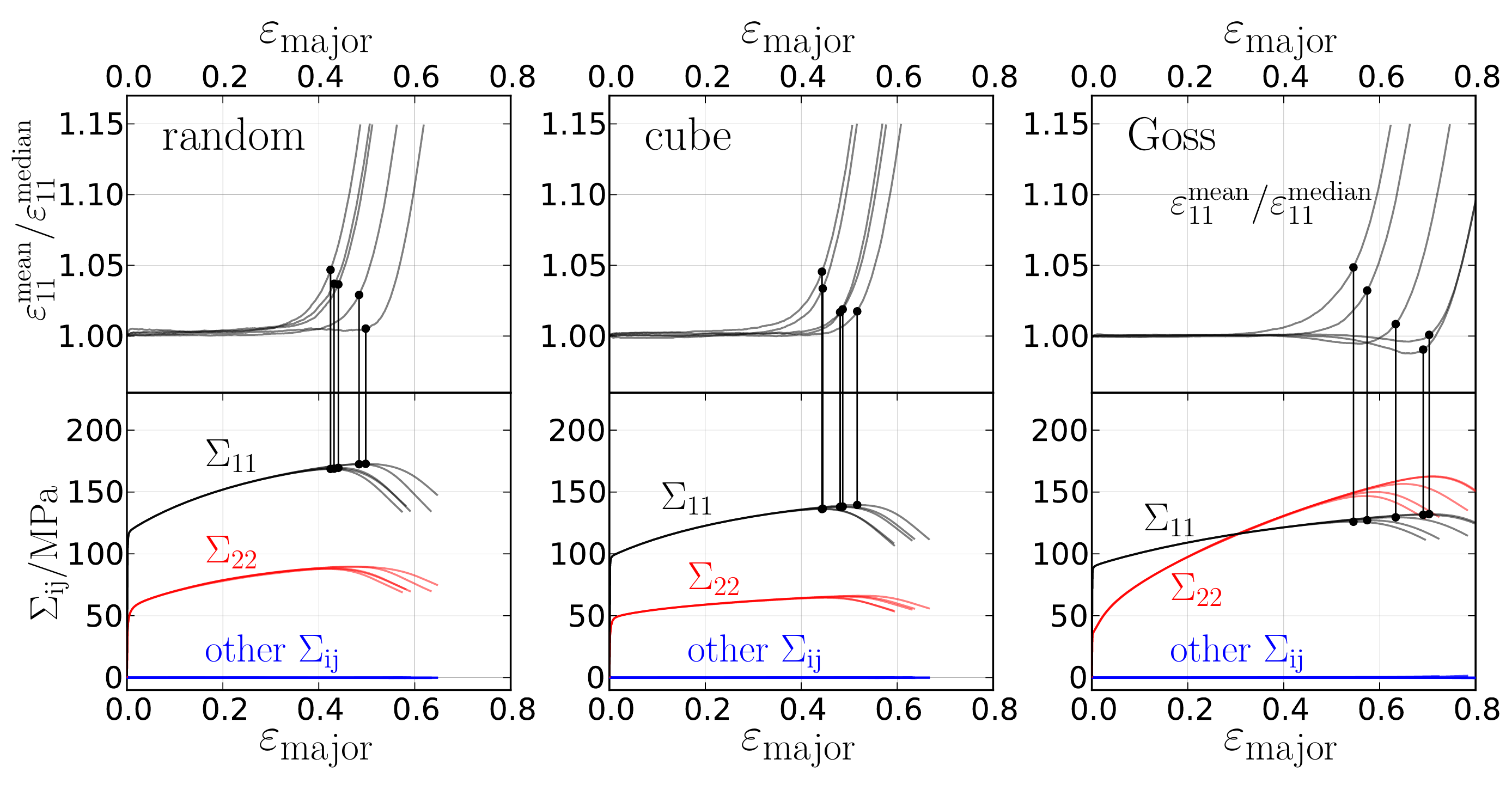}
    \subcaption{$\rho$ = 0.0 (plane strain tension)}
    \label{fig:fl1}
\end{subfigure}
\begin{subfigure}[b]{1.0\linewidth}
    \centering
    \includegraphics[scale=0.55]{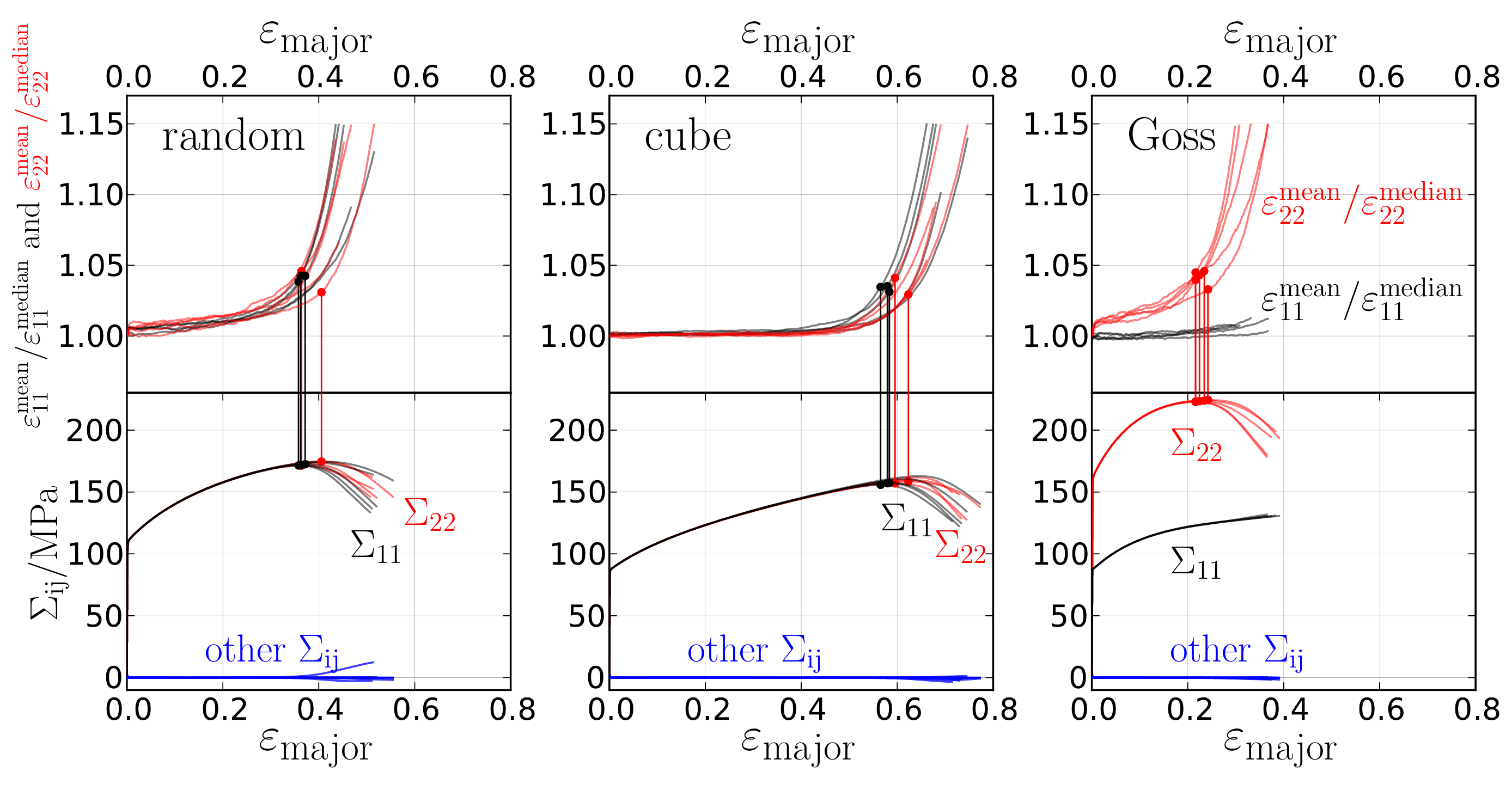}
    \subcaption{$\rho$ = 1.0 (equi-biaxial tension)}
    \label{fig:fl2}
\end{subfigure}
\caption{Top rows show the $\varepsilon_{ii}^{\text{mean}}/\varepsilon_{ii}^{\text{median}}$ ($i$=1,or 2) \vs\ \Emajor~for random (right), cube (middle) and Goss (left) orientation distributions at (a) plane strain tension ($\rho$=0.0) and (b) equi-biaxial tension ($\rho$=1.0). The bottom rows show the macroscopic stresses $\Sigma_{ij}$ \vs\ \Emajor~where the solid circles marks the maximum stresses. Each line corresponds to an orientation distribution instance and all simulations were performed with mesh 20$\times$20$\times$5.}
\label{fig:fl}
\end{figure}

The localization evolution ($\varepsilon_{ii}^{\text{mean}}/\varepsilon_{ii}^{\text{median}}$ \vs\ \Emajor) of the orientation distributions, random, cube, and Goss, are shown in \cref{fig:fl} as examples. At the early stage of deformation, the deformation is usually approximately homogenous, except for the cases of random and Goss at $\rho$ = 1.0 where localization takes place from the beginning of the deformation, yet the localization progresses slowly. At some point, the deformation is mainly carried by localization, indicated by a slope change from $\sim$0 to nearly infinitive. The transition from homogeneous deformation to localized deformation is continuous, and this continuous transition is consistent with the experimental observations made in Ref. \cite{Volk2011}. 

Since the transition from homogeneous to localized deformation is continuous, it is difficult to define a forming limit criterion. Hence, we follow Drucker's stability criterion \cite{Hill1958,Drucker1959,Viatkina2005,Gaenser2000} which states that the deformation instability occurs when the strain softening takes place, \ie\, the forming limit is at the maximum stress on the stress-strain curve. 

\cref{fig:fl} shows the macroscopic stress ($\Sigma_{ij}$) \vs\ \Emajor~of random, cube, and Goss, where $\Sigma_{ij}$ is the average over all the integration points. The maximum stresses are in general occurs when $\varepsilon_{ii}^{\text{mean}}/\varepsilon_{ii}^{\text{median}}$ = 1.05$\sim$1.10, implying that even the localization has been already initiated, the strain hardening is still not exhausted in the localized volume so that the deformation is not fully carried by the localized volume.
\subsection{Simulated texture dependence of forming limit diagrams}
\label{sec:results_fld}
\begin{figure*}
\centering
\includegraphics[scale=0.45]{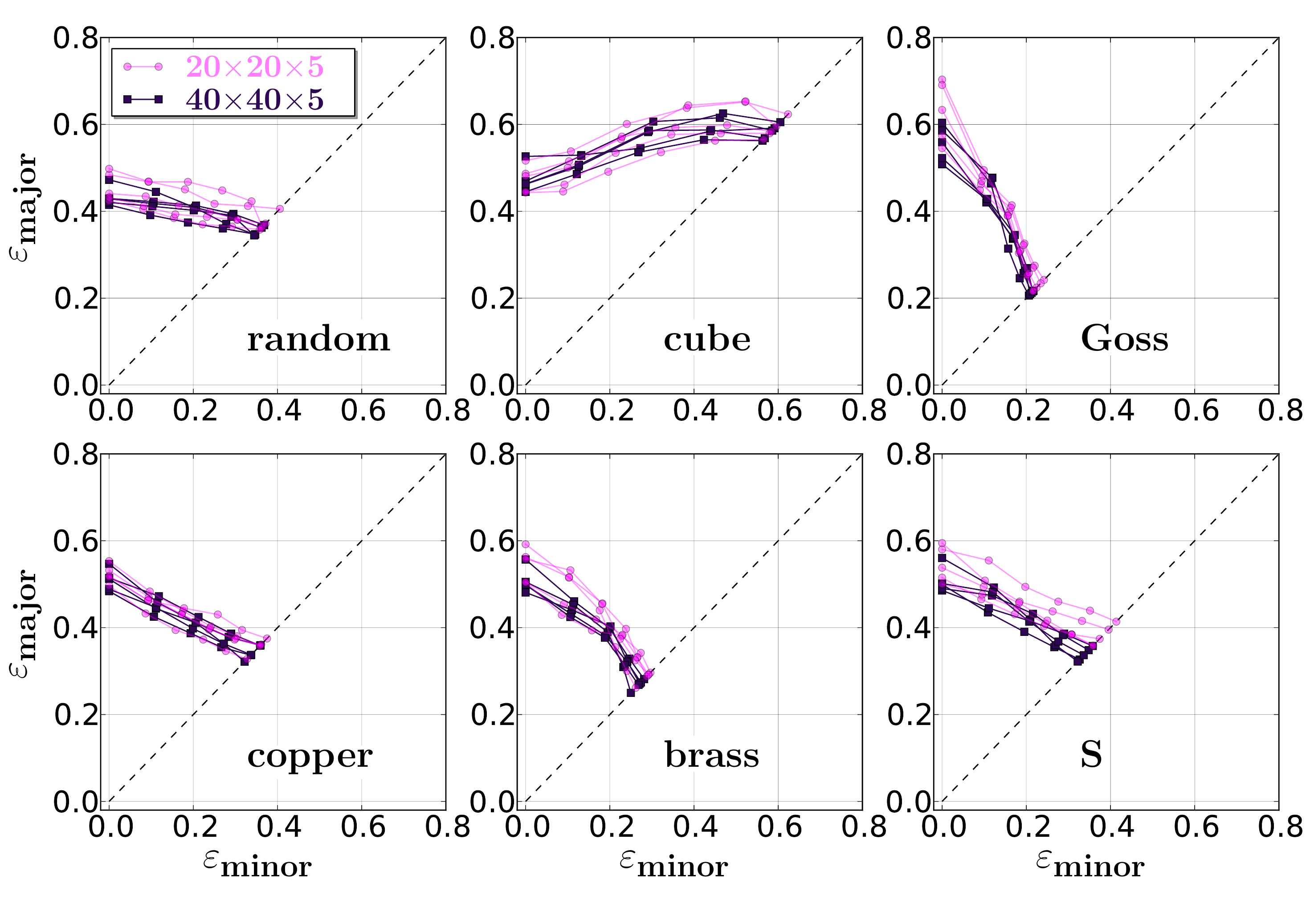}
\caption{Simulated right hand side of FLDs of different orientation distributions. The results of the same orientation distribution instance are connected by lines.}
\label{fig:fld}
\end{figure*}
\begin{figure*}
\centering
\includegraphics[scale=0.45]{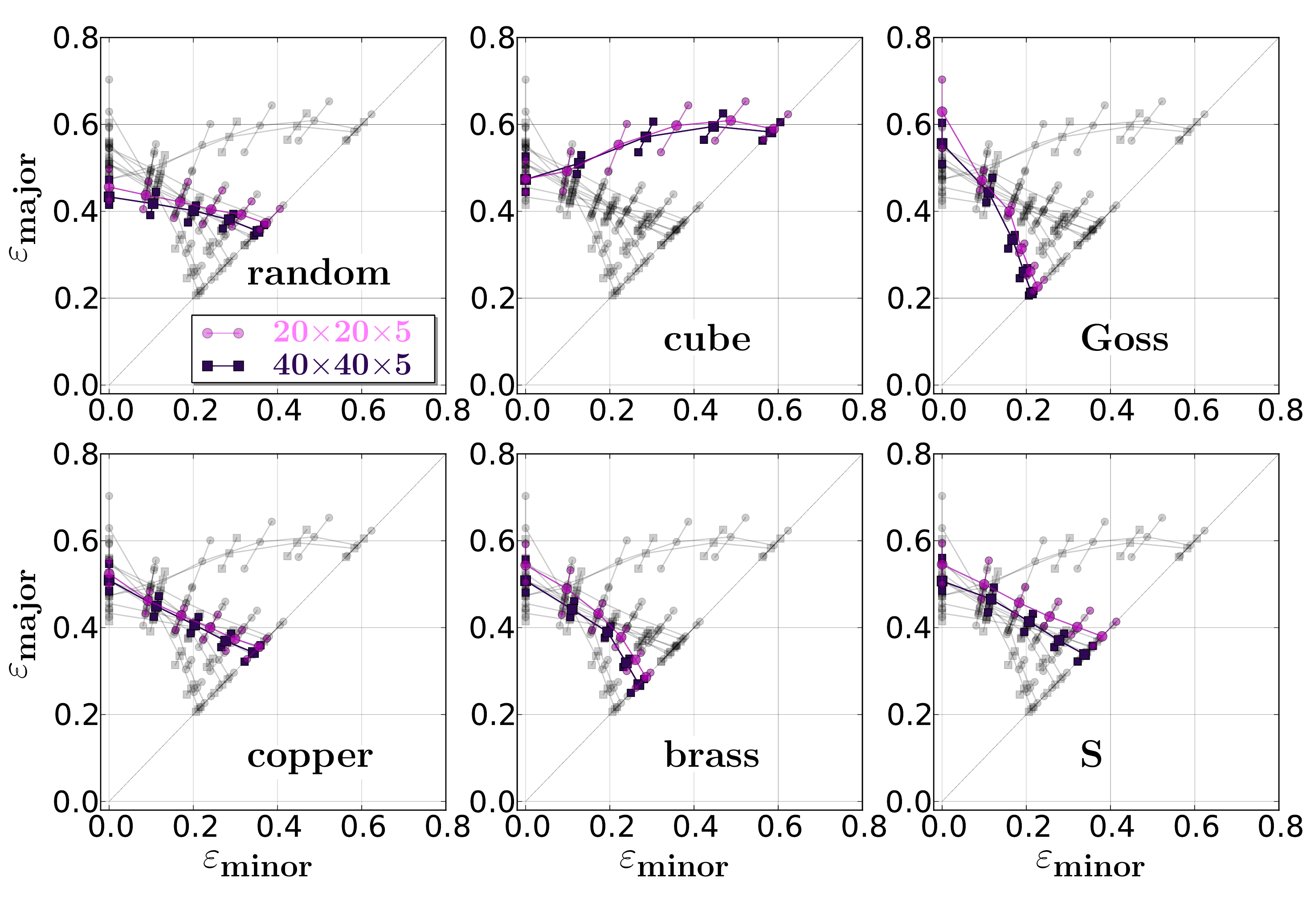}
\caption{Simulated right hand side of FLDs of different orientation distributions. It only shows the mean and extrema values of different orientation distribution instances. All orientation distributions are shown in gray in each figure as a background.}
\label{fig:fld_mean}
\end{figure*}
With the forming limit criterion being defined in the preceding subsection, the simulated FLDs of the texture components are shown in \cref{fig:fld}, and the mean and extrema values are shown in \cref{fig:fld_mean}. 

Though from a statistically point of view, all the orientation distribution instances for a given orientation distribution are the same (see \cref{fig:pf} and the Supplementary Materials), the calculated FLDs still exhibit a certain scatter shown in \cref{fig:fld,fig:fld_mean}. Yet, all the orientation distribution instances still follow the same FLD trend for a given orientation distribution. The scatter introduced by using different distribution instances is not entirely surprising, as the localized deformation is a local event which probably highly depends on the local spatial orientation distribution. 

\cref{fig:fld_mean} also shows the calculated FLDs by using mesh 40$\times$40$\times$5 are slightly lower than those by 20$\times$20$\times$5. A possible reason is the application of PBC. If a localized deformation band is to be formed, the orientation or the shape of the band should fulfill the applied PBC. Thus, PBC may impose certain restriction to the localized deformation. In a larger unit cell, there is more freedom for the localized deformation band to adopt its orientation or shape which is less restricted by the imposed PBC. 

The texture dependence revealed in \cref{fig:fld,fig:fld_mean} is that (i) cube has the best formability, and it is the only orientation distribution whose forming limit increases when $\rho$ increases; (ii) random, copper, and S have comparable formability to each other, and all of them are better than Goss and brass; (iii) Goss has the poorest formability.

The texture dependence in \cref{fig:fld,fig:fld_mean} is actually in good agreement with Fig. 4(f) in Ref. \cite{Yoshida2007}. This agreement is surprising for the main reason that the origin of the localized deformation in this study is the microstructural heterogeneity, while in Ref. \cite{Yoshida2007} (MK-model in conjunction with Taylor model) it is a pre-defined geometrical imperfection. It seems the governing effect is only the crystallograhpic texture. 

This unexpected agreement can be perhaps rationalized as the follows. During the early stage of deformation, small grooves (similar to geometrical imperfection) on the surface develop due to the heterogeneous deformation at the grain scale. The geometrical orientations and the depths of these grooves can be random. Only some of them are able to grow eventually to a volume of localized deformation which carries the overall deformation. 

It should be pointed out some inconsistencies regarding the FLDs of random and cube texture in the literature. For random texture, the trend in this study shown \cref{fig:fld,fig:fld_mean} agrees with the trend in Refs. \cite{Yoshida2007,Signorelli2009} (MK-model with Taylor model) that \Emajor~at $\rho=1.0$ (equi-biaxial tension) is slightly lower than that at $\rho=0.0$ (plane strain tension). With the same method, it shows the \Emajor~at $\rho=1.0$ is almost the same as that at $\rho=0.0$ in Ref. \cite{Wu2004,McGinty2004}. By using another polycrystal homogenization method, \ie\ VPSC with MK-model, it reveals that \Emajor~at $\rho=1.0$ is higher than that at $\rho=0.0$ \cite{Signorelli2009,Signorelli2009a}. For cube texture with 15$^{\circ}$ scatter, the simulated formability in this study is better than that of random, which agrees with simulations by using MK-model with Taylor model \cite{Yoshida2007,Wu2004,Signorelli2009}. Yet, the random is shown to exhibits better formability than cube when MK-model is applied in conjunction with VPSC \cite{Signorelli2009}. 

In Ref. \cite{Signorelli2009}, these inconsistencies are attributed to the texture evolution by using different homogenization models (Taylor model and VPSC). For aluminium alloys rolling texture of predictions, it has been shown that in the case of AA1200, Taylor model (full constraint) gives better prediction than VPSC, yet in the case of AA5182, VPSC is only slightly better than Taylor model \cite{vanHoutte2004}. On the other hand, CP-FEM in general gives better prediction than Taylor model and VPSC \cite{vanHoutte2004,vanHoutte2005}. Hence, these inconsistencies might not be fully attributed to the texture evolution by using different polycrystalline models. 
\section{Simulated FLD of AA6111-T4C in comparison with experiment in Ref. \cite{Wu1998}}
\label{sec:wu1998}
In this section, we simulate the FLD of an aluminium alloy 6111-T4C, and compare the simulation with the experiment in Ref. \cite{Wu1998}. 
\subsection{Simulation input: material parameters and orientation distributions}
\begin{figure}[h!]
\begin{subfigure}[b]{0.5\linewidth}
    \centering
    \includegraphics[scale=0.8]{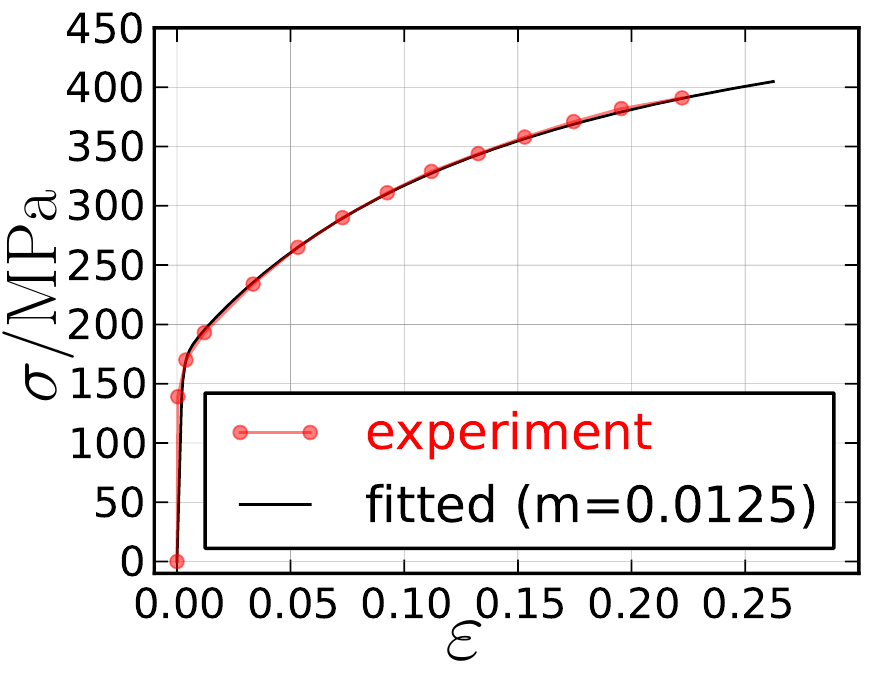}
    \subcaption{parameter set 1 in \cref{tab:wu1998_parameters}}
    \label{fig:wu1998_se1}
\end{subfigure}
\begin{subfigure}[b]{0.5\linewidth}
    \centering
    \includegraphics[scale=0.8]{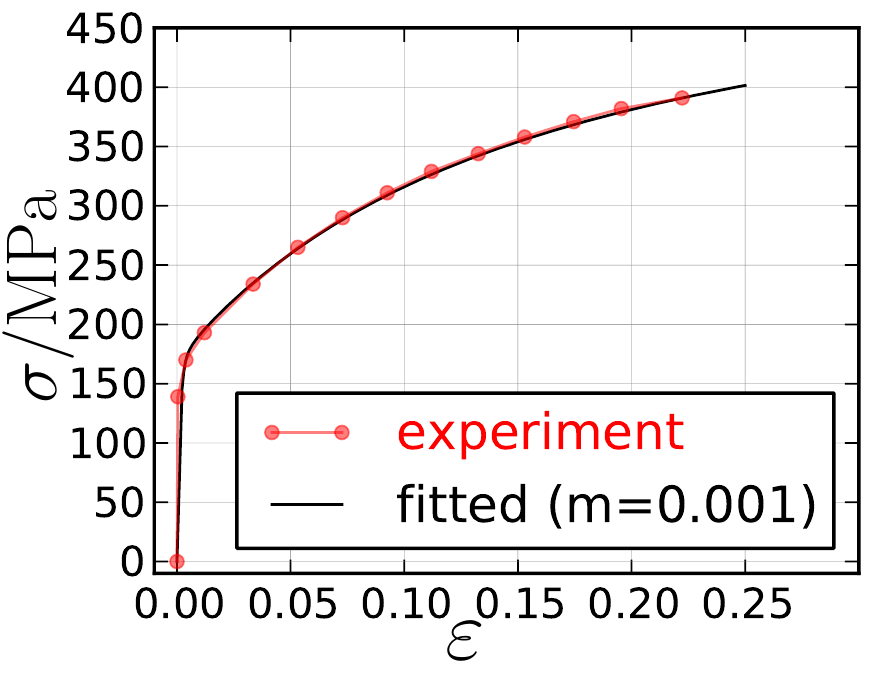}
    \subcaption{parameter set 2 in \cref{tab:wu1998_parameters}}
    \label{fig:wu1998_se2}
\end{subfigure}
\caption{Comparison of the stress-strain curves of AA6111-T4C between the experiment in Ref. \cite{Wu1998} and CP-FEM by using (a) the parameter set 1 and (b) the parameter set 2 in \cref{tab:wu1998_parameters}.}
\label{fig:wu1998_se}
\end{figure}

The materials parameters of AA6111-T4C are obtained by fitting the tensile stress-strain curve (Fig.3d in Ref. \cite{Wu1998}). Two sets of materials parameters are obtained and listed in \cref{tab:wu1998_parameters}. The simulated tensile stress-strain curves together with the experiment are shown in \cref{fig:wu1998_se}. The main difference between those two sets of parameters is the strain rate sensitivity, \ie\, $m=0.0125$ for parameter set 1 and $m=0.001$ for parameter set 2. For both sets, the initial hardening ($h_{0}$) parameters are similar, and the initial critical resolved shear stresses ($\tau^{\alpha}_{c0}$) and saturation shear stresses ($\tau^{\alpha}_{\infty}$) are the same.

\begin{table}[h]
\centering
\caption{Two sets of the material parameters in \cref{eq: hardening pheno,eq: shear rate pheno,eq: sPK} by fitting the stress-strain curve of AA6111-T4C (Fig. 3d in Ref. \cite{Wu1998}). Note $\alpha=1,\ldots,12$. $\tau^{\alpha}_{c0}$ is the initial shear resistance of the slip system $\alpha$. $^{\ast}$ if $\alpha$ and $\beta$ are coplanar slip systems. $^{\ast\ast}$ if $\alpha$ and $\beta$ are not coplanar slip systems.}
\label{tab:wu1998_parameters}
\begin{tabular}{lll|cr|crl}
\toprule
\multicolumn{8}{c}{parameter set 1}\\
\hline
 $\tau^{\alpha}_{c0}$     & 62   & MPa & $\dot{\gamma_{0}}$ & 0.001 & $\tnsrfour{C}_{11}$ & 106.75 & GPa\\
 $\tau^{\alpha}_{\infty}$ & 152  & MPa & $\mathbf{m}$                & {\bf 0.0125}& $\tnsrfour{C}_{12}$ & 60.41  & GPa\\
 $w$                      & 1.2  &     &                    &       & $\tnsrfour{C}_{44}$ & 28.34  & GPa\\
 $h_{0}$ & 412 & MPa & & & & &\\
 $h_{\alpha\beta}$ & 1.0$^{\ast}$ & & & & & &\\
                   & 1.4$^{\ast\ast}$ & & & & & &\\
\hline
\multicolumn{8}{c}{parameter set 2}\\
\hline
 $\tau^{\alpha}_{c0}$     & 62   & MPa & $\dot{\gamma_{0}}$ & 0.001 & $\tnsrfour{C}_{11}$ & 106.75 & GPa\\
 $\tau^{\alpha}_{\infty}$ & 152  & MPa & $\mathbf{m}$               & {\bf 0.001}  & $\tnsrfour{C}_{12}$ & 60.41  & GPa\\
 $w$                      & 1.2  &     &                    &       & $\tnsrfour{C}_{44}$ & 28.34  & GPa\\
 $h_{0}$ & 400 & MPa & & & & &\\
 $h_{\alpha\beta}$ & 1.0$^{\ast}$ & & & & & &\\
                   & 1.4$^{\ast\ast}$ & & & & & &\\									
\bottomrule
\end{tabular}
\end{table}

\begin{figure}[h!]
\centering
\includegraphics[scale=0.3]{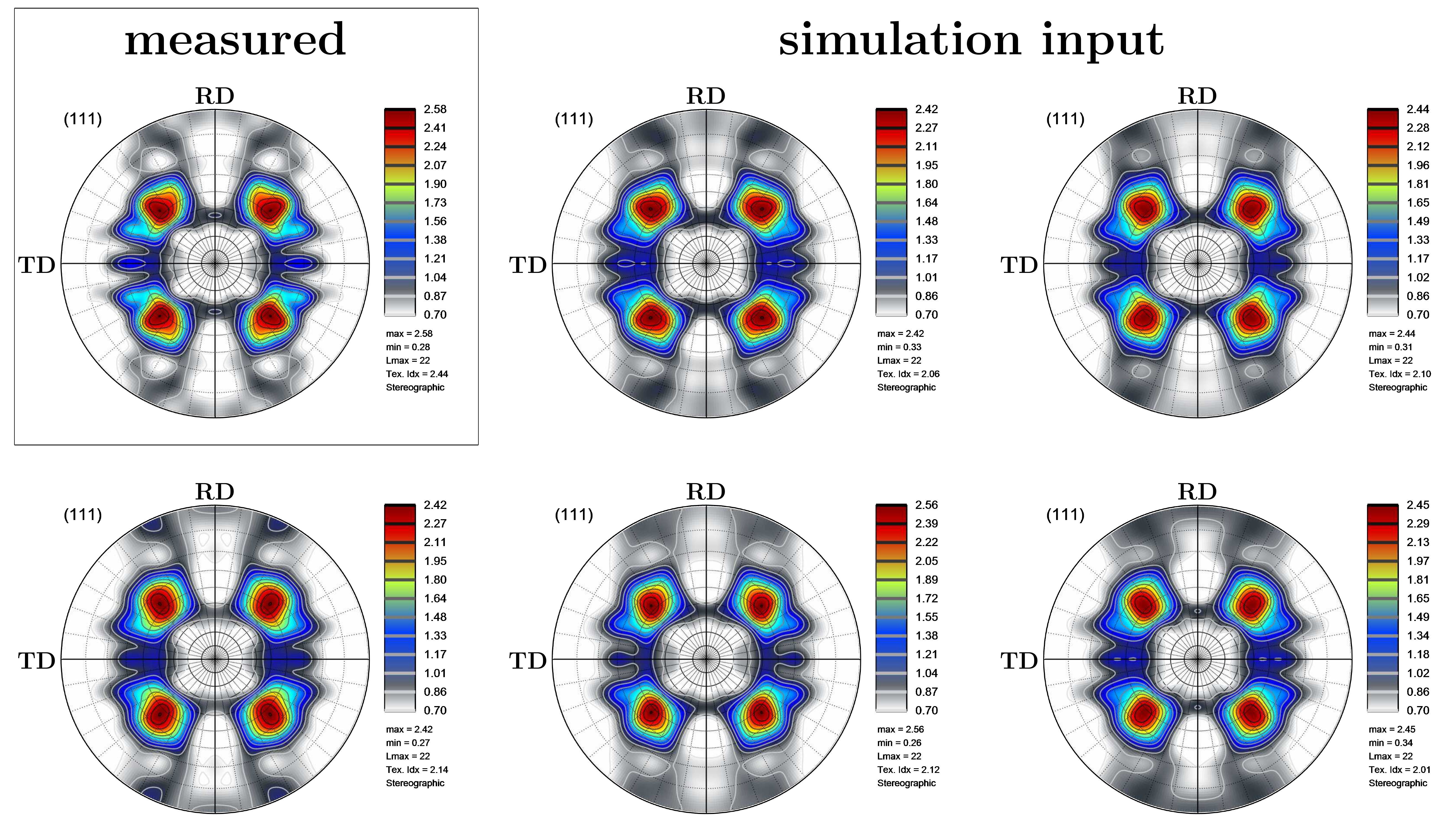}
\caption{Measured pole figure of AA6111-T4C in Ref. \cite{Wu1998} and the pole figures of 5 orientation distribution instances used as simulation input. The orientation distribution simulation input is generated by using the method presented in Ref. \cite{Eisenlohr2008}, and the pole figures are created by using ATEX \cite{jtex}.}
\label{fig:wu1998_texture}
\end{figure}

The measured crystallographic texture of AA6111-T4C from Refs. \cite{Wu2005,Wu1998} is shown in \cref{fig:wu1998_texture}. The measured texture exhibits a typical fcc recrystallization texture, \ie\, approximately 20$\sim$30 \% cube texture with the rest being random texture. The orientation distribution input is generated by using the method presented in Ref. \cite{Eisenlohr2008}. Again 5 orientation distributions instances are created as the simulation input, and their pole figures are also shown in \cref{fig:wu1998_texture}. The simulations were conducted only with the mesh 20$\times$20$\times$5 with or without PBC.

\subsection{Comparison between simulated and measured FLDs of AA6111-T4C}

\begin{figure}[h!]
\centering
\begin{subfigure}[b]{\linewidth}
    \centering
    \includegraphics[scale=0.5]{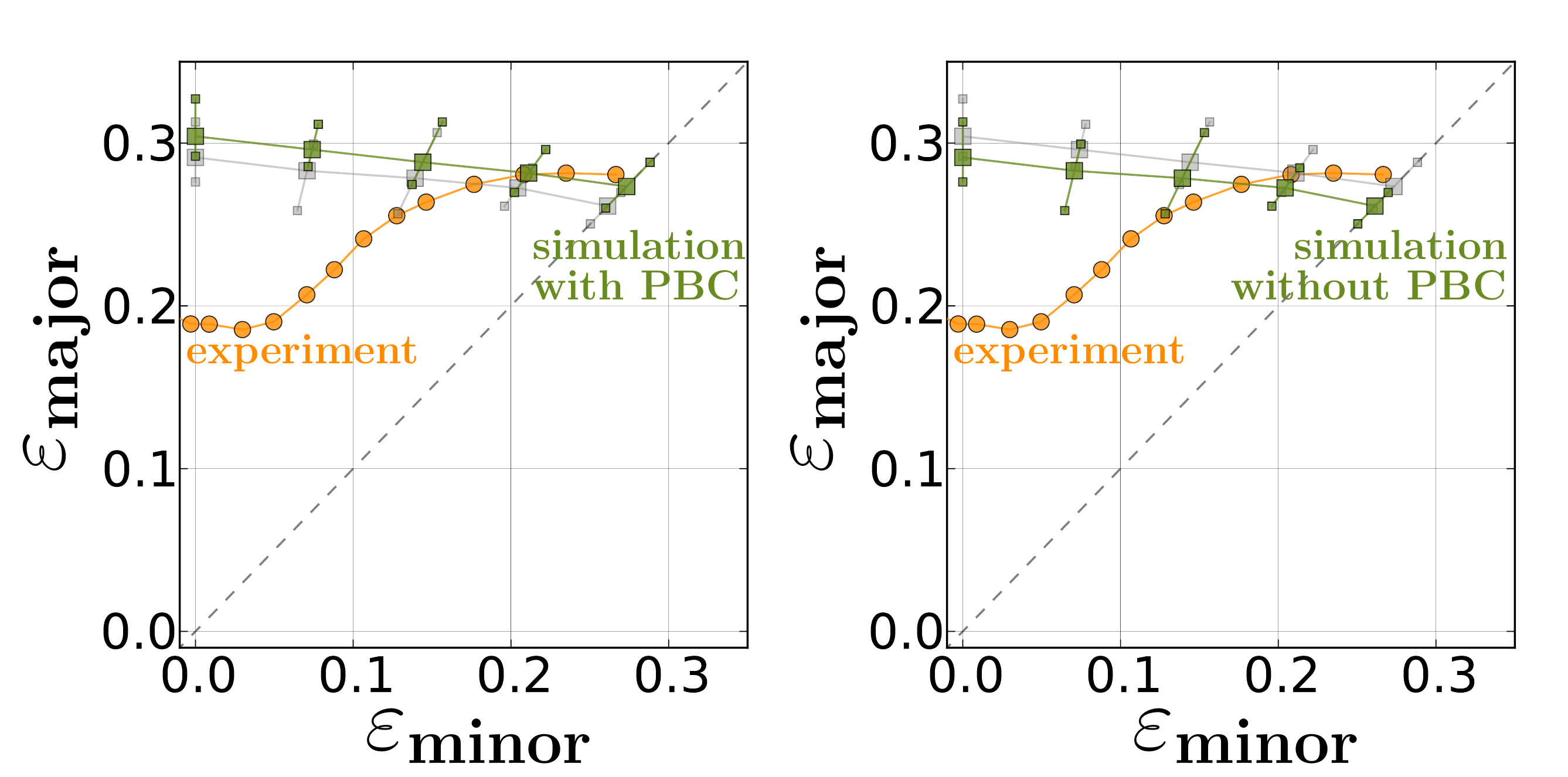}
    \subcaption{parameter set 1 (high strain rate sensitivity) in \cref{tab:wu1998_parameters}}
    \label{fig:wu1998_fld1}
\end{subfigure}
\begin{subfigure}[b]{\linewidth}
    \centering
    \includegraphics[scale=0.5]{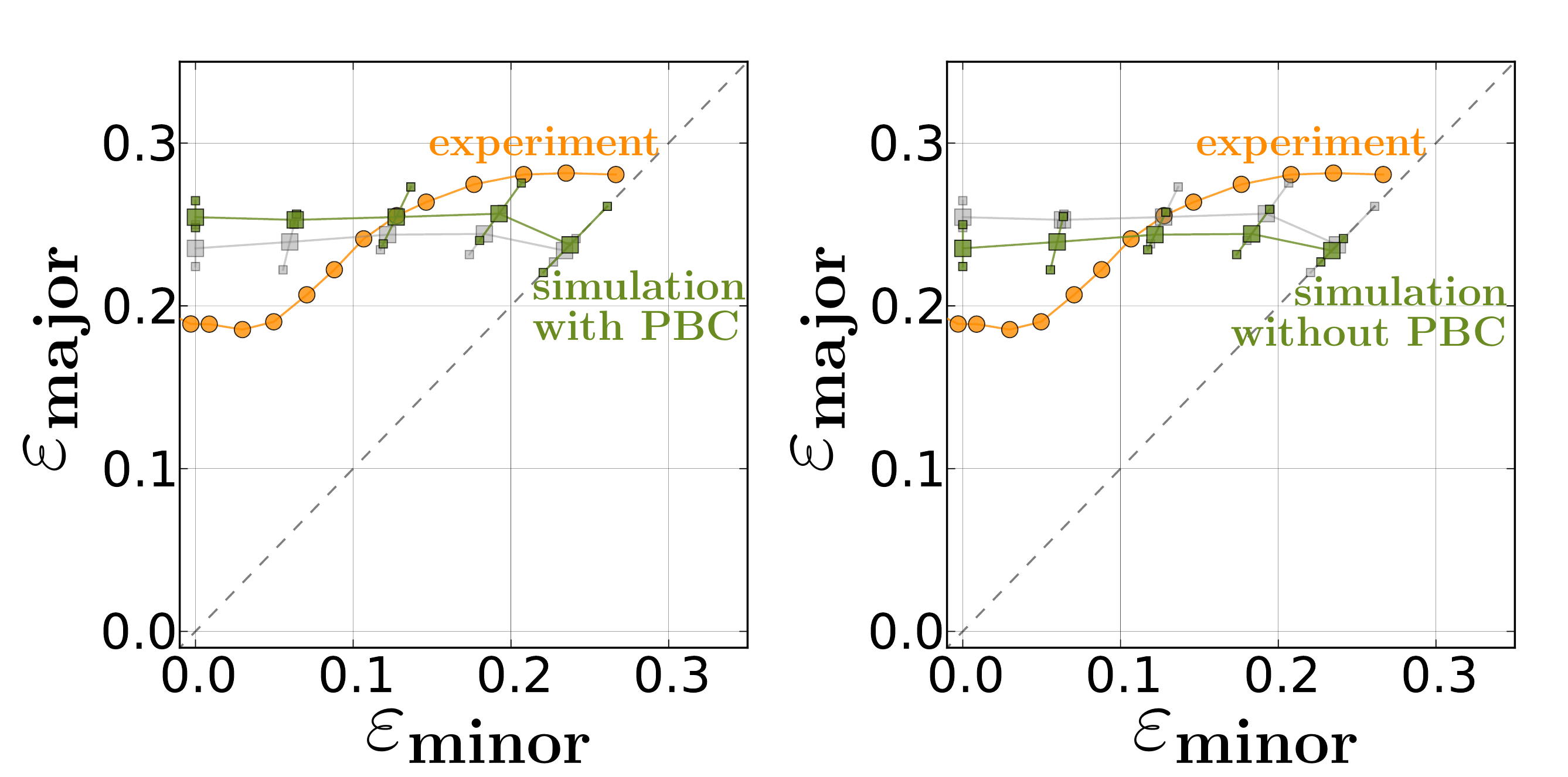}
    \subcaption{parameter set 2 (low strain rate sensitivity) in \cref{tab:wu1998_parameters}}
    \label{fig:wu1998_fld2}
\end{subfigure}
\caption{Comparison between measured (solid orange circles) and simulated (solid green squares) FLDs by using (a) parameter set 1 and (b) parameter set 2 in \cref{tab:wu1998_parameters}. It only shows the mean and extrema values of different orientation distribution instances. Simulations with PBC are on the left hand side and the simulations without PBC are on the right hand side in (a) and (b). In each figure, the simulations with the other boundary condition are shown as gray squares.}
\label{fig:wu1998_fld}
\end{figure}

With the simulation input described in the preceding subsection, the simulated FLDs of AA6111-T4C are shown in \cref{fig:wu1998_fld}. 
It is expected that a higher strain rate sensitivity (parameter set 1 in \cref{tab:wu1998_parameters}) would lead to a better formability than that of a lower strain rate sensitivity (parameter set 2 in \cref{tab:wu1998_parameters}). While the trend of the simulated FLDs by using these two sets of parameters are quite different.

With parameter set 1, \Emajor~at $\rho=0.0$ (plane strain tension) is slightly higher than that at $\rho=1.0$ (equi-biaxial tension), and this trend is similar to the trend shown in \cref{fig:fld,fig:fld_mean} in \cref{sec:yoshida2007} for random texture. This is probably due to two reasons: (1) though the orientation distribution exhibits cube texture (see \cref{fig:wu1998_texture}), there is still $\sim$70 \% volume fraction of random texture; (2) the strain rate sensitivity of parameter set 1 ($m$ = 0.0125) is at the same order of magnitude as that used for \cref{fig:fld,fig:fld_mean} ($m$ = 0.05).

The simulated forming limits without PBC is lower than those with PBC. The difference between the FLDs with and without PBC is nonetheless not significant, and the general trend of the simulated FLDs is not changed.

Among the employed materials parameters and imposed boundary conditions, a better agreement between simulation and experiment can be achieved by using parameter set 2 (low strain rate sensitivity) without PBC. The agreement is, however, only qualitative that \Emajor~slightly increases from $\rho=0.0$ up to $\rho=0.75$, but drops at $\rho=1.0$. Quantitatively, the simulated \Emajor~ at the forming limit at $\rho=0.0$ is higher than the experiment by $\sim$20 \%, and at $\rho=1.0$ it is lower than the experiment by $\sim$20 \%. 

There are some limitations within this study, and they are discussed in the next section which might help with improving the quantitative prediction.
\section{Limitations of this study}
\label{sec:limitation}
Here we discuss three major limitations in this study.

The first limitation is the absence of ductile damage formulation in the constitutive description. In an experimental study on IF (interstitial free) and DP (dual phase) steel \cite{Tasan2009}, it shows that in IF steel damage is initiated by deformation localization due to the lack of damage nucleation sources. While in DP steel, the damage initiates and accumulates before localization where martensite acts as damage nucleation sites. Hence, ductile damage plays a significant role in localization in DP steel. In aluminium alloys, the unwanted but inevitable coarse intermetallic particles often act as damage nucleation sites \cite{Lassance2007}. Thus, it is reasonable to extrapolate from the study in Ref. \cite{Tasan2009} that ductile damage also plays a significant role in the localized deformation in aluminium alloys.

The second limitation is that the number of through-thickness orientations/grains is fixed to be 10, and the grain size is to be 100 $\mu m$. This assumed grain size is at the limit of the rule of thumb that the surface roughness is not visible after forming operation if the grain size is below $\sim$100 $\mu m$ \cite{Yamada1960}. Varying the number of through-thickness grains is, however, not considered in this study. In a simulation study where the localization of columnar grains under plane strain condition (2D model) is investigated \cite{Yoshida2014}, it shows that the limiting strain increases as the number of through-thickness grains increases, and the limiting strain saturates when the number of through-thickness grains exceeds a certain number. We conjecture that the forming limit decreases if the through-thickness grains are less than 10, yet it is not clear how this affects the forming limits on different strain paths.

The third limitation is FEM itself, as it requires considerable computational power. Though in this study DAMASK is implemented in such way that it is compatible with the share memory parallel version of LS-DYNA, and each simulation could be finished within a reasonable time frame, it still requires a lot of computational resources. Hence, it is cost prohibitive to address the influence of some simulation parameters, such as the number of though-thickness grains, the representative volume size, and the in-grain mesh. An better alternative would be the spectral method based solver \cite{Eisenlohr2013,Shanthraj2015}, particularly given the recent demonstration of its capability of including free surface in the model \cite{Maiti2018}. 
\section{Summary}
\label{sec:summary}
We demonstrate that it is feasible to apply 3D CP-FEM models to simulated FLDs without assuming any initial geometrical imperfection. The intrinsic microstructral heterogeneity in polycrystalline model is sufficient to trigger the localization event in the simulations. 

The localized deformation is monitored by the statistical distributions of the principle strains. The observed localization is connected to the stress softening effect, thus the maximum stress appeared on the stress-strain curve is proposed as the criterion to identify the forming limit. 

FLDs are simulated for 6 typical fcc texture components. The simulated texture dependence is in good agreement with a previous simulation study by Yoshida \textit{et al.} \cite{Yoshida2007} where the MK-model in conjunction with Taylor model is employed. Though based on different physical origins of localization, it is argued that the surface roughness or grooves induced by the early stage plastic deformation of polycrystalline materials could serve as the geometrical imperfection assumed in the MK-model. Thus, the crystallographic texture becomes the dominant effect on the resulting FLD.

FLD of AA6111-T4C is also simulated with a real texture as input, and the simulated FLD is compared with the experiment by Wu \textit{et al.} \cite{Wu1998}. Only a qualitative agreement can be achieved within this study. The simulated \Emajor~of the forming limit at $\rho=0.0$ (plane strain tension) is higher than the experiment by $\sim$20 \%, and at $\rho=1.0$ (equi-biaxial tension) it is lower than the experiment by $\sim$20 \%. The limitations within this study are discussed which might be subjected to future studies and improve the quantitative prediction.

\section*{Acknowledgments}
The author would like to thank Dr. Martin Diehl and PD Dr. Franz Roters at Max-Planck-Institut f\"{u}r Eisenforschung, D\"{u}sseldorf, Germany for the initial discussion on incorporating DAMASK and LS-DYNA. The author would also like to express his gratitude to Dr. Tobias Erhart at DYNAmore GmbH, Stuttgart, Germany for the discussion of user defined subroutines in LS-DYNA. 
Prof. Peidong Wu at McMaster University, Canada is very much appreciated for his generosity of providing the author with the experimental data. 
This work has been supported by the European Regional Development Fund (EFRE) in the framework of the EU-program "IWB Investition in Wachstum und Besch\"{a}ftigung \"{O}sterreich 2014-2020", and the federal state Upper Austria.
\section*{Reference}
\providecommand{\newblock}{}

\end{document}